\documentclass[a4paper,12pt]{article}
\pdfoutput=1
\usepackage{amsfonts,amsthm,amsmath,amssymb,graphicx,hyperref,color,youngtab}
\def\ket#1{\left| #1\right\rangle}

\def\Tr{{\rm Tr\, }}

\newcommand{\be}{\begin{equation}}
\newcommand{\bea}{\begin{eqnarray}}

\newcommand{\ee}{\end{equation}}
\newcommand{\eea}{\end{eqnarray}}

\begin{document}

\makeatletter
\@addtoreset{equation}{section}
\makeatother
\renewcommand{\theequation}{\thesection.\arabic{equation}}

\rightline{}%WITS-CTP-108
\vspace{1.8truecm}

\vspace{15pt}

%%%%%%%%%%%%%%%%%

{\LARGE{  
\centerline{\bf Interacting Double Coset Magnons} 
}}  

\vskip.5cm 

\thispagestyle{empty} 
\centerline{{\large \bf 
Abdelhamid Mohamed Adam Ali\footnote{\tt  1134940@students.wits.ac.za}
Robert de Mello Koch\footnote{ {\tt robert@neo.phys.wits.ac.za}}}, }
\centerline{\large \bf Nirina Hasina Tahiridimbisoa\footnote{\tt NirinaMaurice.HasinaTahiridimbisoa@students.wits.ac.za}
and  Augustine Larweh Mahu\footnote{ {\tt 1130820@students.wits.ac.za} }}

\vspace{.4cm}
\centerline{{\it National Institute for Theoretical Physics ,}}
\centerline{{\it School of Physics and Mandelstam Institute for Theoretical Physics,}}
\centerline{{\it University of Witwatersrand, Wits, 2050, } }
\centerline{{\it South Africa } }

\vspace{1.4truecm}

%%%%%%%%%%%%%%%%%
\thispagestyle{empty}

\centerline{\bf ABSTRACT}

\vskip.4cm 
We consider the anomalous dimensions of restricted Schur polynomials constructed using $n\sim O(N)$ complex adjoint scalars
$Z$ and $m$ complex adjoint scalars $Y$. 
We fix $m\ll n$ so that our operators are almost half BPS. 
At leading order in ${m\over n}$ this system corresponds to a dilute gas of $m$ free magnons.
Adding the first correction of order ${m\over n}$ to the anomalous dimension, which arises at two loops, 
we find non-zero magnon interactions.
The form of this new operator mixing is studied in detail for a system of two giant gravitons with four strings attached.

\setcounter{page}{0}
\setcounter{tocdepth}{2}

\newpage

\tableofcontents

\setcounter{footnote}{0}

\linespread{1.1}
\parskip 4pt

{}~
{}~

\section{Introduction}

The original instance of the AdS / CFT correspondence\cite{malda} provides a definition for a class of quantum type IIB string
theories: those that are embedded in spacetimes which are asymptotically $AdS_5\times S^5$ with background five form flux. 
The definition for this class of string theories is in terms of the highly symmetric superconformal four dimensional 
${\cal N} = 4$ super Yang-Mills theory with gauge group $U(N)$. 
The correspondence claims a one-to-one and onto mapping between states of the quantum gravity and quantum 
operators of the gauge theory. 
Consequently, the mapping will identify all the objects known to string theory, perturbative and non-perturbative, with
operators in ${\cal N} = 4$ super Yang-Mills theory. 
Apart from the perturbative spectrum of closed strings, there are D-branes and their open string excitations, as well 
as other spacetime geometries, living in the gauge theory.

An interesting class of D-branes are the giant graviton branes\cite{mst,myers,hash}.
We now have a good idea of how to describe the operators that correspond to certain examples of these branes.
The examples we have in mind are almost ${1\over 2}$-BPS giant gravitons.
The dual operators are built from two complex scalar fields $Z$ and $Y$ of the ${\cal N}=4$ super Yang-Mills theory.
We need to use order $N$ fields, so that these operators have a large dimension in the large $N$ limit.
For operators with such a large classical dimension the usual large $N$ techniques, i.e. an expansion
organized by genus of contributing ribbon graphs, is not possible\cite{bbns}.
New techniques to study the large $N$ limit have been developed.
For the free theory bases of operators that diagonalize the two point function to all orders in $1/N$ have been 
identified\cite{cjr1,cjr2,dssi,Kimura:2007wy,BHR1,BHR2,Bhattacharyya:2008rb,Kimura:2008ac,Kimura:2009jf,Kimura:2012hp,Pasukonis:2013ts}.
Techniques to study the anomalous dimension of these operatotrs have also been developed, first for descriptions
in which the giant graviton plus open string system is treated using words in the gauge theory to represent the open 
string\cite{Balasubramanian:2002sa,Berenstein:2003ah,Balasubramanian:2004nb,Berenstein:2005fa,Berenstein:2006qk,dssii,bds} 
and secondly for descriptions which treat all 
fields in the operator democratically\cite{Carlson:2011hy,Koch:2011hb,gs,deMelloKoch:2012ck}.
Our operators are built using $n\sim O(N)$ $Z$s and $m$ $Y$s with $n\gg m$, which implies that we are close to the 
${1\over 2}$-BPS giant graviton and that we have a new small parameter ${m\over n}$ in the game.
 The condition that $n\gg m$ is crucuial for our approximations which is not too surprising: a systematically small
deformation of a BPS operator will be simpler than the generic operator.
Part of the motivation for this paper is to consider the first order correction in a systematic ${m\over n}$ expansion.

In both the open string description and in the more democratic description, there is a close 
connection\cite{chrisnirina} between
the dynamics of the $Y$ fields, and the LLM plane\cite{LLM} description 
of giant magnon dynamics in the dual string theory\cite{db,hm1,hm2}. 
In this study we are interested in the anomalous dimensions of gauge theory operators corresponding to restricted 
Schur polynomials that treat the fields democratically. 
This corresponds to a dilute gas of magnons.
The spectrum of anomalous dimensions has been computed to all loops at large $N$ and to leading order in the small 
parameter ${m\over n}$\cite{stuart,chrisnirina}.
This all loop answer is possible thanks to supersymmetry\cite{Beisert:2005tm}.
The result agrees with explicit one loop computations in\cite{Koch:2011hb,deMelloKoch:2012ck},
two loop computations in \cite{twoloop}
and even three loop computations performed in \cite{threeloop}
using a collective coordinate\cite{Berenstein:2013md,Berenstein:2013eya,Berenstein:2014isa,Berenstein:2014zxa}
approach.
The result has an interesting structure which is worth understanding to appreciate the results we report here.
The operator we interested in is defined by
\bea
   \chi_{R,(r,s)\alpha\beta}(Z,Y)={1\over n!m!}\sum_{\sigma\in S_{n+m}}
\chi_{R,(r,s)\alpha\beta}(\sigma){\rm Tr}(\sigma Z^{\otimes n}\otimes Y^{\otimes m})
\eea
The labels $R\vdash n+m$, $r\vdash n$, $s\vdash m$ label irreducible representations of $S_{n+m}$, $S_n$ and $S_m$
respectively.
$(r,s)$ labels a representation of the subgroup $S_n\times S_m$.
The above polynomial is only non-zero if $(r,s)$ arises from $R$ after restricting to the $S_n\times S_m$ subgroup.
Since it may arise more than once, we need multiplicity labels (denoted $\alpha$ and $\beta$ above) to keep track of the
copy we are considering.
One way to ensure that $(r,s)$ arises after restricting to the subgroup, is to realize $r$ by removing $m$ boxes from Young
diagram $R$.
These removed boxes are then reassembled to give $s$.
Use $m_i$ to count the number of boxes that must be removed from row $i$ of $R$ to get $r$.
Assemble the $m_i$ to produce the vector $\vec{m}$.
The vector $\vec{m}$ is conserved to leading order in ${m\over n}$.

In this article we will study the first subleading corrections in ${m\over n}$ to the anomalous dimension.
This explores the first contributions which induce magnon interactions for the dilute magnon gas.
If we are ever to understand the non-perturbative sectors of string theory using the gauge theory / gravity 
correspondence, it seems that we must move beyond small deformations of ${1\over 2}$-BPS operators.
One way to do this is to construct a good understanding of this system, beyond the leading order in ${m\over n}$.
This is a key motivation for this paper.

There are already corrections of order ${m\over n}$ to the one loop anomalous dimension.
These corrections do not lead to new operator mixing - they only correct the anomalous dimension.
The first non-trivial corrections appear at two loops.
Consequently, in section 2 we review the structure of the two loop dilatation operator.
It is useful to rewrite the action of the two loop dilatation operator in a basis, the Gauss graph basis,
that diagonalizes the leading order.
In this way it will become apparent that the subleading correction induces new operator mixing.
In the process of transforming to the Gauss graph basis we encounter new types of traces that have not
been computed before.
In the Appendix \ref{howtotrace} we develop techniques powerful enough to evaluate the most general
trace we could encounter, which is much more general than the traces that appear at two loops.
These results will be useful in further studies of the dynamics of Gauss graph operators.
To illustrate our results, in section \ref{2branes4strings} we consider a state of two giant gravitons with 
four strings attached. Our results show a number of interesting features. Operators labeled by different
Gauss graphs start to mix implying that we do indeed have an interacting system of magnons.
As a consequence of these interactions, the vector $\vec{m}$ is no longer conserved and operators
with different $\vec{m}$ labels mix. 
Finally, at the leading order in ${m\over n}$ and at large $N$, the all loop dilatation operator factorizes 
into an action on the $Y$s times an action on the $Z$s.
Although this factorization was only exhibited at one loop in \cite{Koch:2011hb} and at two loops in \cite{twoloop}, 
the arguments of \cite{stuart} as well as the form of the all loop anomalous dimension \cite{Beisert:2005tm}
implies that this factorization holds to all loops; see \cite{chrisnirina} for further discussion.
The subleading term that we have evaluated does not factorize into an action on the $Y$s times an action on the $Z$s.
This proves that the action of the dilatation operator only factorizes into an action on the $Y$s times an action on the $Z$s
at the leading order in a systematic ${m\over n}$ expansion.
In section \ref{discussion} we discuss these results and suggest a number of interesting directions in which the present 
study can be extended.

\section{The Two Loop Dilatation Operator}

The complete two loop dilatation operator in the $su(2)$ sector is given by\cite{Beisert:2003tq}
\bea
D_4=D_4^{(1)}+D_4^{(2)}+D_4^{(3)}
\eea
where
\bea
  D_4^{(1)} &=&-2 g^2:{\rm Tr}\left(\left[ \left[Y,Z\right],{\partial\over\partial Z}\right]
       \left[ \left[{\partial\over\partial Y},{\partial\over\partial Z}\right], Z\right]\right):\cr
  D_4^{(2)} &=&-2 g^2:{\rm Tr}\left(\left[ \left[Y,Z\right],{\partial\over\partial Y}\right]
       \left[ \left[{\partial\over\partial Y},{\partial\over\partial Z}\right], Y\right]\right):\cr
  D_4^{(3)} &=&-2 g^2:{\rm Tr}\left(\left[ \left[Y,Z\right],T^a \right]
        \left[ \left[{\partial\over\partial Y},{\partial\over\partial Z}\right], T^a \right]\right):
\label{dilop}
\eea
and
\bea
g={g_{YM}^2\over 16\pi^2}
\eea
The sum over $a$ in $D_4^{(3)}$ is easily performed with the help of the identity
\bea
{\rm Tr}(T^a A T^a B)={\rm Tr}(A){\rm Tr}(B)\label{uncomp}
\eea
which follows from the completeness of the $T^a\in u(N)$.
The size of these three terms is easily estimated as follows: $D_4^{(1)}$ has two derivatives with respect to $Z$
which act on $n$ $Z$ fields and one with respect to $Y$ which acts on $m$ $Y$ fields.
The size of this term is thus $\sim n^2 m$.
Similarly, we estimate that $D^{(2)}\sim n m^2$ and $D^{(3)}\sim Nnm\sim n^2 m$. 
Thus, the leading order comes from $D_4^{(1)}$ and $D_4^{(3)}$, while the first $O({m\over n})$ correction to the
leading term comes from $D_4^{(2)}$.
In section \ref{lead} we will review the results for the action of $D_4^{(1)}$ and $D_4^{(3)}$.
In the process we will introduce the basis of operators, the Gauss graph operators, that diagonalizes the action 
of the leading dilatation operator.
Following this, we study the action of $D^{(2)}_4$ in section \ref{sublead}.
To transform this action to the Gauss graph basis requires that we develop new techniques to evaluate certain
traces that appear.
These techniques are developed in Appendix \ref{howtotrace} and the action of $D^{(2)}_4$ in the Gauss graph
basis is discussed in section \ref{sublead}.

\subsection{Leading Contribution}\label{lead}

Acting on a restricted Schur polynomial, the action of $D_4^{(1)}$ is
\bea
  D_4^{(1)}\chi_{R,(r,s)\alpha\beta}&=& g^2 \sum_{T,(t,u)\mu\nu} \Big(
 L^{(a)}_{{{T,(t,u)\mu\nu}};{R,(r,s)\gamma\delta}}
+L^{(b)}_{{{T,(t,u)\mu\nu}};{R,(r,s)\gamma\delta}}\Big) \chi_{T,(t,u)\gamma\delta}
\eea
where
\bea
 L^{(a)}_{{{T,(t,u)\mu\nu}};{R,(r,s)\gamma\delta}} 
&=&-\sum_{R''\, ,\, T''} {d_T n (n-1)m\over d_t d_u d_{R''}(n+m)(n+m-1)}
c_{R,R'}c_{R',R''}\Bigg[\cr
&& {\rm Tr}\Big(I_{T''\, R''}(2,m+2,m+1) P_{R,(r,s)\alpha\beta} [(1,m+2)-(1,m+1)] \cr
&&\,\times (2,m+2) I_{R''\, T''}  (2,m+2) P_{T,(t,u)\delta\gamma} (2,m+2) [(m+1,2,1)-(2,m+1,1)]\Big)\cr
&+& {\rm Tr}\Big(I_{T''\, R''}(2,m+2)[(1,m+1)- (m+2,m+1)] P_{R,(r,s)\alpha\beta} (1,m+2,2) I_{R''\, T''}\cr
&&\times \big(
(1,m+1) (2,m+2) P_{T,(t,u)\delta\gamma} (2,m+2)(2,m+1)\cr
&&-(m+1,2,m+2) P_{T,(t,u)\delta\gamma} (2,m+2)(2,1)\big)\Big) \Bigg]
\eea
and
\bea
 L^{(b)}_{{{T,(t,u)\mu\nu}};{R,(r,s)\gamma\delta}}
&=&\sum_{R'\, ,\, T'} {d_T n(n-1)m\over d_t d_u d_{R'} (n+m)}c_{R,R'}\Bigg[\cr
&&\times {\rm Tr}\Big(I_{T'\, R'}[(1,m+2,m+1)P_{R,(r,s)\alpha\beta} - (m+2,m+1)P_{R,(r,s)\alpha\beta}(1,m+1)]I_{R'\, T'}\cr
&&\times 
[(1,m+1)P_{T,(t,u)\delta\gamma}-P_{T,(t,u)\delta\gamma}(1,m+1)]\Big)\cr
&&+ {\rm Tr}\Big(I_{T'\, R'}[(1,m+1,m+2)P_{R,(r,s)\alpha\beta} - (m+2,m+1)P_{R,(r,s)\alpha\beta}(1,m+2)]I_{R'\, T'}\cr
&&\times 
[(1,m+1)P_{T,(t,u)\delta\gamma}-P_{T,(t,u)\delta\gamma}(1,m+1)]\Big)\Bigg]
\eea
In the above expression the traces run over the direct sum of the carrier spaces  $R\oplus T$.
The Young diagrams $R$ and $T$ both label irreducible representations of $S_{n+m}$.
Primes denote Young diagrams obtained by dropping boxes, with one box dropped for each prime.
Thus, for example, $T''$ is an irreducible representation of $S_{n+m-2}$, obtained by dropping two
boxes from $T$. 
The factors $I_{T'\, R'}$ and $I_{T''\, R''}$ are intertwining maps mapping from the carrier space
$T'$ to $R'$ and from $T''$ to $R''$ respectively.
$c_{RR'}$ is the factor of the box that must be dropped from $R$ to get $R'$.
We use a little letter to denote dimensions of irreducible representations of the symmetric group so that, for example,
$d_R$ is the dimension of the symmetric group representation labeled by Young diagram $R$.
Finally, $P_{R,(r,s)\alpha\beta}$ denotes the intertwining maps which correctly restrict the trace
in $R$ to the subspace relevant for the restricted character, that is
$\chi_{R,(r,s)\alpha\beta}(\sigma)={\rm Tr}(P_{R,(r,s)\alpha\beta}\Gamma^{(R)}(\sigma))$.

The above result is exact in the sense that all orders in $1/N$ are included.
The traces appearing in the above expression run over the direct sum of carrier spaces $R\oplus T$.
To exploit the simplifications of the large $N$ limit, we now employ the distant corners approximation.
In this approximation, the traces over $R\oplus T$ are reduced to a trace over the tensor product of
the direct sum of the carrier spaces $r\oplus t$ and $V_p^{\otimes m}$ where $R$ has a total of $p$ rows
and $V_p$ is a $p$ dimensional vector space.
The trace over $r\oplus t$ is rather straight forward.
The bulk of the work then entails tracing over $V_p^{\otimes m}$.
From now on we will work with normalized restricted Schur polynomials $O_{R,(r,s)\alpha\beta}$ which
are scaled version of the $\chi_{R,(r,s)\alpha\beta}$
\bea
O_{R,(r,s)\alpha\beta}(Z,Y)=\sqrt{{\rm hooks}_r{\rm hooks}_s\over f_R {\rm hooks}_R}\chi_{R,(r,s)\alpha\beta}(Z,Y)
\eea
To denote the length of a given row of a Young diagram, we will indicate the Young diagram label with a subscript 
which identifies the row.
Thus $r_1$ is the length of the first row of $r$ and $T_2$ is the length of the second row in $T$.
After tracing over $r\oplus t$, we have
\bea
&&D_4^{(1)}O_{R,(r,s)\alpha\beta}(Z,Y)=
-2g^2\sum_{i<j}{m\over\sqrt{d_u d_s}}
\Big({\rm Tr}_{V_p^{\otimes m}}
(E^{(1)}_{ii}p_{R,(r,s)\alpha\beta}E^{(1)}_{jj}p_{T,(t,u)\delta\gamma})\cr
&&\qquad\qquad+{\rm Tr}_{V_p^{\otimes m}}
(E^{(1)}_{jj}p_{R,(r,s)\alpha\beta}E^{(1)}_{ii}p_{T,(t,u)\delta\gamma})\Big)
\Delta_{ij}O_{T,(t,u)\gamma\delta}(Z,Y)\label{leading}
\eea
where
\bea
&&\Delta_{ij}O_{T,(t,u)\gamma\delta}(Z,Y)\equiv \Big(
\sqrt{(N+R_2)(N+R_2-1)(N+R_1-1)(N+R_1-2)}
\delta_{T''_{jj},R''_{ii}}\delta_{t''_{jj},r''_{ii}}\cr
&&+\sqrt{(N+R_2-2)(N+R_2-3)(N+R_1)(N+R_1+1)}\delta_{T''_{ii},R''_{jj}}\delta_{t''_{ii},r''_{jj}}\cr
&&-(2N-3)\left(\sqrt{(N+R_1-1)(N+R_2-1)}\delta_{T'_j,R'_i}\delta_{t'_j,r'_i}
+\sqrt{(N+R_2-2)(N+R_1)}\delta_{T'_i,R'_j}\delta_{t'_i,r'_j}\right)\cr
&&+[2(N+R_1-1)(N+R_2-2)-(n-1)(2N+n-3)]\delta_{T,R}\delta_{t,r}
\Big)O_{T,(t,u)\gamma\delta}(Z,Y)\cr
&&
\eea
and of course $n=r_1+r_2$.
The delta functions which appear are 1 if the Young diagram labels have the same shape and 0 otherwise.
In the above formula, the matrices $E^{(A)}_{ij}$ which appear are a basis for the representation of $u(p)$ on
$V_p^{\otimes m}$. 
Concretely, $E_{ij}$ is a matrix with every entry equal to zero except for the entry in the $i$th row and $j$th column,
which is equal to 1.
In terms of $E_{ij}$ we can write
\bea
   E^{(A)}_{ij}={\bf 1}\otimes {\bf 1}\otimes\cdots \otimes E_{ij}\otimes\cdots \otimes {\bf 1}
\eea
where $E_{ij}$ appears in the $A$th factor of the tensor product and ${\bf 1}$ is the $p$ dimensional unit matrix.
To obtain the result (\ref{leading}) we have used
\bea
\sqrt{{\rm hooks}_T{\rm hooks}_r\over{\rm hooks}_R{\rm hooks}_t}=1\left( 1+O\left({m\over n}\right)\right)
\label{sublead}
\eea
and we have only kept the leading order. 
The action of $D_4^{(1)}$ is a product of two factors: an action on Young diagrams $R,r$ and an independent action
on Young diagram $s$.
To evaluate this second action explicitely, we need to trace over $V_p^{\otimes m}$.
This is most easily achieved by moving to the basis of Gauss graph operators.
Each Gauss graph operator is labeled by an element of the double coset $H\setminus S_m/H$ where 
$H=S_{m_1}\times S_{m_2}\times \cdots\times S_{m_p}$.
The relation between the Gauss graph operator ($O_{R,r}(\sigma)$) and the normalized restricted
Schur polynomial is
\bea
  O_{R,r}(\sigma)
  ={|H|\over \sqrt{m!}}\sum_{j,k}\sum_{s\vdash m}\sum_{\mu_1,\mu_2}\sqrt{d_s}
  \Gamma^{(s)}_{jk}(\sigma )B^{s\to 1_H}_{j \mu_1}B^{s\to 1_H}_{k \mu_2} O_{R,(r,s)\mu_1\mu_2}
  \label{ggo}
\eea
This transformation can be understood as a Fourier transform applied to the double coset.
The branching coefficients $B^{s\to 1_H}_{j \mu_1}$ give a resolution of the projector from the irreducible 
representation $s$ of $S_m$ to the trivial representation of $H$
\bea
{1\over |H|}\sum_{ \sigma \in H } \Gamma^{(s)}_{ik} ( \sigma )
=\sum_\mu B^{s \rightarrow 1_H}_{ i \mu}  B^{s \rightarrow 1_H}_{ k \mu}
\eea
In terms of the Gauss graph operator, we find
\bea\label{actiondil} 
   D^{(1)}_4 O_{R,r}(\sigma) = -2g^2 \sum_{i<j}n_{ij} ( \sigma ) \Delta_{ij}\, O_{R,r}(\sigma)
\eea
The numbers $n_{ij}(\sigma)$ can be read off of the element of the double coset element $\sigma$.
For more details see Appendix \ref{DCoset}

After using (\ref{uncomp}), the action of $D_4^{(3)}$ reduces to the action of the one loop dilatation operator.
Consequently, we will not dicsuss this term further.

\subsection{Subleading Contribution}\label{sublead}

There are a number of different sources for the subleading contribution. 
Firstly, the leading two loop terms computed above recieve corrections - see equation (\ref{sublead}).
These corrections do not lead to additional mixing.
They only imply a correction to the anomalous dimension.
Similarly, even the one loop anomalous dimension recieves an ${m\over n}$ correction (also from using the approximation
given in (\ref{sublead})), without any additional operator mixing.
The first correction that implies new operator mixing comes from the leading contribution to $D_4^{(2)}$ and it will be the 
focus of this subsection.
This term has not been considered before.

Acting on a normalized restricted Schur polynomial, we find
\begin{eqnarray}
D_4^{(2)}O_{R,(r,s)\alpha\beta}(Z,Y) &=&  \sum_{T,(t,u)\mu\nu} \Big(
 M^{(a)}_{{{T,(t,u)\mu\nu}};{R,(r,s)\alpha\beta}}
-M^{(b)}_{{{T,(t,u)\mu\nu}};{R,(r,s)\alpha\beta}}
-M^{(c)}_{{{T,(t,u)\mu\nu}};{R,(r,s)\alpha\beta}}\cr
&+&M^{(d)}_{{{T,(t,u)\mu\nu}};{R,(r,s)\alpha\beta}}\Big)O_{T,(t,u)\mu\nu}(Z,Y),
\end{eqnarray}
where
\begin{eqnarray}
M^{(a)}_{{{T,(t,u)\mu\nu}};{R,(r,s)\alpha\beta}} &=& 
\sum_{R',R''}
{\frac{d_T nm(m-1) c_{RR'}c_{R'R''} }{d_t d_u (n+m)(n+m-1)d_{R''}}}
\sqrt{f_T{\rm hooks}_T{\rm hooks}_r{\rm hooks}_s\over f_R{\rm hooks}_R{\rm hooks}_t{\rm hooks}_u}
\nonumber\\
&&
\times\Tr \Big(I_{T''R''}\Gamma_R((2,m+1))
\left[ \Gamma_R((1,m+1)) , P_{R,(r,s)\alpha\beta}\right] I_{R''T''}\nonumber\\
&&\times \left(
\Gamma_T((1,m+1))\left[\Gamma_T((2,m+1)), P_{T,(t,u)\mu\nu}\right] \right)\Big),\nonumber
\end{eqnarray}
\begin{eqnarray}
M^{(b)}_{{{T,(t,u)\mu\nu}};{R,(r,s)\alpha\beta}} &=& 
\sum_{R'}\frac{d_T nm(m-1)c_{RR'}}{d_t d_u (n+m) d_{R'}} 
\sqrt{f_T{\rm hooks}_T{\rm hooks}_r{\rm hooks}_s\over f_R{\rm hooks}_R{\rm hooks}_t{\rm hooks}_u}
\nonumber\\
&&\times\Tr\Big( \Gamma_R((2,m+1))
\left[ \Gamma_R((1,m+1)) , P_{R,(r,s)\alpha\beta}\right] I_{R'T'}
\nonumber\\
&&\times
\left[ \Gamma_T((1,m+1)), P_{T,(t,u)\mu\nu}\right] I_{T'R'} \Big),\nonumber
\end{eqnarray}
\begin{eqnarray}
M^{(c)}_{{{T,(t,u)\mu\nu}};{R,(r,s)\alpha\beta}} &=& 
\sum_{R'}\frac{d_T nm(m-1)c_{RR'}}{d_t d_u (n+m) d_{R'}} 
\sqrt{f_T{\rm hooks}_T{\rm hooks}_r{\rm hooks}_s\over f_R{\rm hooks}_R{\rm hooks}_t{\rm hooks}_u}
\nonumber\\
&&\times\Tr  \Big( \left[ \Gamma_R((1,m+1)), P_{R,(r,s)\alpha\beta}\right]\Gamma_R((2,m+1))
I_{R'T'}\nonumber\\
&&\times  \left[ \Gamma_T((1,m+1)) ,P_{T,(t,u)\mu\nu}\right]  I_{T',R'} \Big),\nonumber
\end{eqnarray}
and
\begin{eqnarray}
M^{(d)}_{{{T,(t,u)\mu\nu}};{R,(r,s)\alpha\beta}}&=&
\sum_{R',R''}
{\frac{d_T nm(m-1) c_{RR'}c_{R'R''} }{d_t d_u (n+m)(n+m-1)d_{R''}}}
\sqrt{f_T{\rm hooks}_T{\rm hooks}_r{\rm hooks}_s\over f_R{\rm hooks}_R{\rm hooks}_t{\rm hooks}_u}
\nonumber\\
&&\times\Tr\Big( I_{T''R''} \left[ \Gamma_R((1,m+1)), P_{R,(r,s)\alpha\beta}\right]\Gamma_R((2,m+1))\nonumber\\
&&\times I_{R''T''}
\left[\Gamma_T((1,m+1)), P_{T,(t,u)\mu\nu}\right] \Gamma_T((2,m+1))
 \Big).\nonumber
\end{eqnarray}
The traces appearing above again run over the direct sum of carrier spaces $R\oplus T$ and
the action given above is again correct to all orders in $1/N$.
To take advantage of the simplifications of the large $N$ limit, we again employ the distant corners approximation,
which again leads to an expression that has a single trace over $V_p^{\otimes m}$ remaining
\begin{eqnarray}
M^{(a)}_{{{T,(t,u)\mu\nu}};{R,(r,s)\alpha\beta}} 
&=&  \sum_{R'R''} \delta_{R''T''}\frac{d_T m(m-1)n c_{RR'}c_{R'R''}}
{d_{R''} d_t d_u (m+n)(m+n-1)}
\sqrt{\frac{f_T{\rm hooks}_T{\rm hooks}_r{\rm hooks}_s}
{f_R{\rm hooks}_R{\rm hooks}_t{\rm hooks}_u}}\,  d_{r^\prime_i}  \delta_{r^\prime_i t^\prime_l}\nonumber\\
&&\times\Bigg[\Tr_{V_p^{\otimes m}}
\left( E^{(1)}_{kj}E^{(2)}_{ll}p_{s\alpha\beta}E^{(1)}_{ii}E^{(2)}_{jk}p_{u\mu\nu}\right) 
-\Tr_{V_p^{\otimes m}}
\left(E^{(1)}_{kj}p_{s\alpha\beta}E^{(1)}_{ii}E^{(2)}_{jk}p_{u\mu\nu}\right)\,\delta_{kl}\nonumber\\
&&\quad
-\Tr_{V_p^{\otimes m}}\left( E^{(1)}_{ki}E^{(2)}_{ll}p_{s\alpha\beta}E^{(2)}_{ik}p_{u\mu\nu}\right)\,\delta_{ij}
+\Tr_{V_p^{\otimes m}}\left(E^{(1)}_{ki}p_{s\alpha\beta}E^{(2)}_{ik}p_{u\mu\nu}\right)\,\delta_{ij}\delta_{kl} \Bigg]
\nonumber
\end{eqnarray}
\begin{eqnarray}
M^{(b)}_{{{T,(t,u)\mu\nu}};{R,(r,s)\alpha\beta}} 
&=&  \sum_{R'} \delta_{R'T'}\frac{d_T m(m-1)n c_{RR'}}{d_{R'} d_t d_u (m+n)}
\sqrt{\frac{f_T{\rm hooks}_T{\rm hooks}_r{\rm hooks}_s}
{f_R{\rm hooks}_R{\rm hooks}_t{\rm hooks}_u}}\, d_{r^\prime_i}  \delta_{r^\prime_i t^\prime_k} \nonumber \\
&&\quad\times\Bigg[\Tr_{V_p^{\otimes m}}\left( E^{(1)}_{ka}E^{(2)}_{ak}p_{s\alpha\beta}
E^{(1)}_{ii}p_{u\mu\nu}\right)  -\Tr_{V_p^{\otimes m}}
\left(E^{(1)}_{ba}E^{(2)}_{ab}p_{s\alpha\beta}E^{(1)}_{ii}p_{u\mu\nu}\right)\, \delta_{ik}\nonumber\\
&&\qquad-\Tr_{V_p^{\otimes m}}\left( p_{u\mu\nu} E^{(1)}_{ki}E^{(2)}_{ik}p_{s\alpha\beta}\right) 
+\Tr_{V_p^{\otimes m}}\left(E^{(1)}_{bi}E^{(2)}_{ib}p_{s\alpha\beta}E^{(1)}_{kk}p_{u\mu\nu}\right)\Bigg],\nonumber
\end{eqnarray}
\begin{eqnarray}
M^{(c)}_{{{T,(t,u)\mu\nu}};{R,(r,s)\alpha\beta}} 
&=&  \sum_{R'} \delta_{R'T'}\frac{d_T m(m-1)n c_{RR'}}{d_{R'} d_t d_u (m+n)}
\sqrt{\frac{f_T{\rm hooks}_T{\rm hooks}_r{\rm hooks}_s}
{f_R{\rm hooks}_R{\rm hooks}_t{\rm hooks}_u}}\, d_{r^\prime_i}  \delta_{r^\prime_i t^\prime_k}\nonumber \\
&&\times\Bigg[
\Tr_{V_p^{\otimes m}}\left( E^{(1)}_{kk}p_{s\alpha\beta}E^{(1)}_{ic}E^{(2)}_{ci}p_{u\mu\nu}\right)  
-\Tr_{V_p^{\otimes m}}\left(p_{s\alpha\beta}E^{(1)}_{ik}E^{(2)}_{ki}p_{u\mu\nu}\right) \nonumber\\
&&\quad
-\Tr_{V_p^{\otimes m}}\left( E^{(1)}_{ii}p_{s\alpha\beta}E^{(1)}_{ab}E^{(2)}_{ba}p_{u\mu\nu}\right)\, \delta_{ik}  
+\Tr_{V_p^{\otimes m}}\left(E^{(1)}_{ii}p_{s\alpha\beta}E^{(1)}_{ck}E^{(2)}_{kc}p_{u\mu\nu}\right)\Bigg],\nonumber
\end{eqnarray}
and
\begin{eqnarray}
&&
M^{(d)}_{{{T,(t,u)\mu\nu}};{R,(r,s)\alpha\beta}} 
= \sum_{R'R''} \delta_{R''T''}\frac{d_T m(m-1)n c_{RR'}c_{R'R''}}{d_{R''} d_t d_u (m+n)(m+n-1)}
\sqrt{\frac{f_T{\rm hooks}_T{\rm hooks}_r{\rm hooks}_s}
{f_R{\rm hooks}_R{\rm hooks}_t{\rm hooks}_u}}\,d_{r^\prime_i}  \delta_{r^\prime_i t^\prime_k}\nonumber \\
&&\qquad\times\Bigg[
\Tr_{V_p^{\otimes m}}\left( E^{(1)}_{ka}E^{(2)}_{ai}p_{s\alpha\beta}E^{(1)}_{ib}E^{(2)}_{bk}p_{u\mu\nu}\right)
\, \delta_{ij}\delta_{kl}  
-\Tr_{V_p^{\otimes m}}\left(E^{(1)}_{la}E^{(2)}_{ai}p_{s\alpha\beta}E^{(1)}_{ik}E^{(2)}_{kl}p_{u\mu\nu}\right)
\, \delta_{ij} \nonumber\\
&&\qquad
-\Tr_{V_p^{\otimes m}}\left( E^{(1)}_{ki}E^{(2)}_{ij}p_{s\alpha\beta}E^{(1)}_{jb}E^{(2)}_{bk}p_{u\mu\nu}\right)
\, \delta_{lk}  +\Tr_{V_p^{\otimes m}}
\left(E^{(1)}_{li}E^{(2)}_{ij}p_{s\alpha\beta}E^{(1)}_{jk}E^{(2)}_{kl}p_{u\mu\nu}\right) \Bigg].
\nonumber
\end{eqnarray}
To compute the remaining trace over $V_p^{\otimes m}$ we will again move to the Gauss graph basis.
This requires computing traces that have not been considered in previous works. 
Schematically, these traces are of the form
\bea
{\rm Tr}(Ap_{s\mu\nu}Bp_{u\gamma\delta})
\eea
where $A$ and $B$ can be any product of the $E^{(A)}_{ab}$s.
The details of how to compute these traces in general are given in Appendix \ref{howtotrace}.
To summarize the key ideas, consider an intertwining map $p_{s\mu\nu}$ built on the state $|\bar{v}_1,\vec{m}_1\rangle$ 
with symmetry group $H_1$ and the map $p_{t\gamma\delta}$ built on the state $|\bar{v}_2,\vec{m}_2\rangle$ with 
symmetry group $H_2$.
The transformation to Gauss graph basis is given by
\bea
T&=&\sum_{s\,\mu\,\nu\, l\, m}
\sum_{u\,\gamma\,\delta\, n\, p}
{\rm Tr}(Ap_{s\mu\nu}Bp_{u\gamma\delta})
B^{s\to 1_H}_{m\mu}B^{s\to 1_H}_{l\nu}\Gamma^{(s)}_{ml}(\sigma_1 )
B^{u\to 1_H}_{p\gamma}B^{u\to 1_H}_{n\delta}\Gamma^{(u)}_{pn}(\sigma_2 )\cr
&=&{1\over |H_1||H_2|}\sum_{\psi_i\in S_m}
\langle v_2|\sigma_2^{-1}\, (\psi_2^{-1}A\psi_2 )\, \psi_1 |v_1\rangle
\langle v_2|\,(\psi_2^{-1}B^T\psi_2)\,\psi_1\sigma_1 |v_1\rangle
\eea
The final sum over $\psi_1$ and $\psi_2$ then gives
\bea
  T={(m-2)!\over |H_1||H_2|}\, |O(\sigma_1)|^2\, n_{ab}(\sigma_2)\, n_{cd}(\sigma_2)\,
\delta_{AB}\left([\sigma_1],[\sigma_2]\right)
\eea
The indices $a,b,c,d$ are read from $A$ and $B$ as explained in Appendix \ref{howtotrace} and
\bea
|O(\sigma_1)|^2=\prod_{i=1}^p n_{ii}(\sigma_1)!\prod_{k,l=1, l\ne k}^p n_{kl}(\sigma_1)!
=\langle O_{R,r}(\sigma_1)^\dagger O_{R,r}(\sigma_1)\rangle
\eea
The delta function $\delta_{AB}\left([\sigma_1],[\sigma_2]\right)$ is defined using $A$ and $B$.
This is also explained in Appendix \ref{howtotrace}.
Using these results, we find
\bea
D_4^{(2)} O_{R,r}(\sigma_1) = \sum_{T,t,\sigma_2} \left(M^{1,\sigma_1,\sigma_2}_{T,t;R,r} 
- M^{2,\sigma_1,\sigma_2}_{T,t;R,r}- M^{3,\sigma_1,\sigma_2}_{T,t;R,r} + M^{4,\sigma_1,\sigma_2}_{T,t;R,r}\right) 
O_{T,t}(\sigma_2),\label{fnlsbl}
\eea
where
\bea
M^{1,\sigma_1,\sigma_2}_{T,t;R,r}&=& \sum_{R',R''} \delta_{R''T''}\delta_{r^\prime_{i}t^\prime_l}\sqrt{\frac{c_{RR'}c_{R'R''}c_{TT'}c_{T'T''}}{R_jT_k}} \, 
\left| O_{R,r}(\sigma_1)\right|^2 \delta_{AB}
\left([\sigma_1],[\sigma_2]\right) \nonumber\cr
&&\times\bigg[n_{ik}(\sigma_2)n_{kl}(\sigma_2) - \delta_{kl}\sum_{a=1}^p n_{ik}(\sigma_2)n_{ka}(\sigma_2) -\delta_{ij}\sum_{b=1}^{p}n_{bk}(\sigma_2)n_{kl}(\sigma_2)\cr
&&\qquad + \delta_{ij}\delta_{kl}\sum_{a,b=1}^{p} n_{bk}(\sigma_2)n_{ka}(\sigma_2)\bigg],
\eea
Similarly
\bea
M^{2,\sigma_1,\sigma_2}_{T,t;R,r} &=& \sum_{R'} \delta_{R'T'}
\delta_{r^\prime_{i}t^\prime_k}\sqrt{c_{RR'}c_{TT'}} \, \left|O_{R,r}(\sigma_1)\right|^2 
\delta_{AB}\left([\sigma_1],[\sigma_2]\right)  \nonumber\cr
&&\times\bigg[\sum_{a,b=1}^{p} n_{ik}(\sigma_2)n_{ba}(\sigma_2) - \delta_{ik}\sum_{a,b,c=1}^p  n_{ib}(\sigma_2)n_{ca}(\sigma_2)-\sum_{a,b=1}^{p}n_{ak}(\sigma_2)n_{bi}(\sigma_2)\cr
&&\qquad + \sum_{a,b=1}^{p}n_{kb}(\sigma_2)n_{ai}(\sigma_2)\bigg],
\eea
\bea
M^{3,\sigma_1,\sigma_2}_{T,t;R,r}&=&\sum_{R'} \delta_{R'T'}\delta_{r^\prime_{i}t^\prime_k}
\sqrt{c_{RR'}c_{TT'}} \, \left|O_{R,r}(\sigma_1)\right|^2 \delta_{AB}\left([\sigma_1],[\sigma_2]\right)\cr
&&\times\bigg[ \sum_{a,b=1}^{p} n_{bk}(\sigma_2)n_{ia}(\sigma_2) 
- \sum_{a,b=1}^{p}  n_{ka}(\sigma_2)n_{ib}(\sigma_2) 
-\delta_{ik}\sum_{a,b,c=1}^p n_{bi}(\sigma_2)n_{ac}(\sigma_2) \cr
&&\qquad+ \sum_{a,b=1}^{p} n_{ki}(\sigma_2)n_{ba}(\sigma_2)\bigg],
\eea
\bea
M^{4,\sigma_1,\sigma_2}_{T,t;R,r}&=&\sum_{R',R''} \delta_{R''T''}\delta_{r^\prime_{i}t^\prime_k}
\sqrt{\frac{c_{RR'}c_{R'R''}c_{TT'}c_{T'T''}}{R_j T_l }} \, \left|O_{R,r}(\sigma_1)\right|^2 
\delta_{AB}\left([\sigma_1],[\sigma_2]\right)  \cr
&&\times\bigg[\delta_{ij}\delta_{kl}\sum_{a,b=1}^{p} n_{bl}(\sigma_2)n_{la}(\sigma_2) 
- \delta_{ij}\sum_{a=1}^p n_{kl}(\sigma_2)n_{la}(\sigma_2)
-\delta_{kl}\sum_{b=1}^{p} n_{bl}(\sigma_2)n_{li}(\sigma_2)\cr
&&\qquad + n_{kl}(\sigma_2)n_{li}(\sigma_2)\bigg],
\eea
Notice that both $M^{1,\sigma_1,\sigma_2}_{T,t;R,r}$ and $M^{4,\sigma_1,\sigma_2}_{T,t;R,r}$ depend on the length of
the rows of the Young diagrams $R$ and $T$ that participate.
Since these lengths determine the angular momentum of the giants, they determine the radius to which the giants will expand.
This is the first dependence of the anomalous dimensions on the geometry of the giant graviton.

To illustrate the form of operator mixing captured in the above formulas, it is helpful to consider a specific example.
This is the content of the next section.

\section{Example: A 2 Giant Graviton Boundstate with 4 Strings Attached}\label{2branes4strings}

In this section we will consider the simplest nontrivial system that exhibits the general structure of the subleading
operator mixing problem.
This problem is a system of two giant gravitons with four strings attached.
As we have explained, there are ${m\over n}$ corrections to the anomalous dimension from the one loop
contribution as well as from $D^{(1)}_4$ and $ D^{(3)}_4$ at two loops.
These corrections do not induce extra operator mixing, so that the Gauss graph operators continue to have 
a good anomlaous dimension.
The only extra operator mixing comes from the subleading contibution $D^{(2)}_4$, and this is what we aim
to explore.

Each element of the Gauss graph basis is labeled by two Young diagrams $R,r$ as well as a Gauss graph.
We will number the states according to their Gauss graph labeling as shown below
\begin{align*}
&\ket{1,R,r} = \left|\raisebox{-.4\height}{\includegraphics[scale=.55]{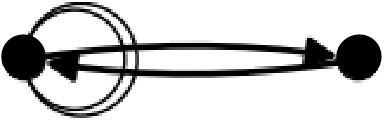}},R,r\right\rangle,
\quad \ket{2,R,r} = \left|\raisebox{-.4\height}{\includegraphics[scale=.55]{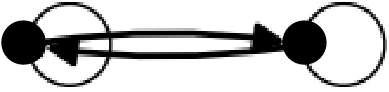}},R,r\right\rangle,
\quad \ket{3,R,r} = \left|\raisebox{-.4\height}{\includegraphics[scale=.55]{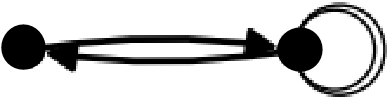}},R,r\right\rangle,\\
&\ket{4,R,r} = \left|\raisebox{-.4\height}{\includegraphics[scale=.55]{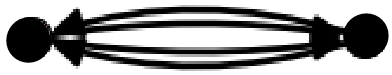}},R,r\right\rangle, 
\quad\ket{5,R,r} = \left|\raisebox{-.4\height}{\includegraphics[scale=.55]{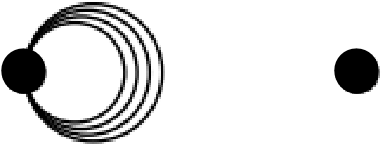}},R,r\right\rangle,
\quad \ket{6,R,r} = \left|\raisebox{-.4\height}{\includegraphics[scale=.55]{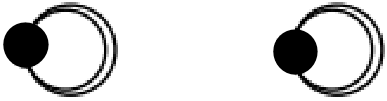}},R,r\right\rangle,\\
& \ket{7,R,r} = \left|\raisebox{-.4\height}{\includegraphics[scale=.55]{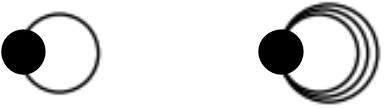}},R,r\right\rangle,
\quad \ket{8,R,r} = \left|\raisebox{-.4\height}{\includegraphics[scale=.55]{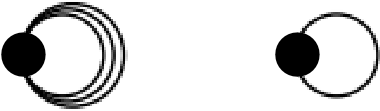}},R,r\right\rangle, 
\quad\ket{9,R,r} = \left|\raisebox{-.4\height}{\includegraphics[scale=.55]{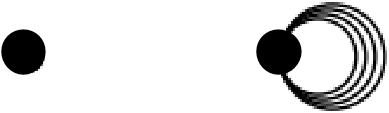}},R,r\right\rangle.
\end{align*}
The states $\ket{i}$ for $i=5,6,7,8,9$ are BPS at the leading order in ${m\over n}$.
The subleading corrections to the anomalous dimension coming from one loop, as well as from $D^{(1)}_4$ and 
$ D^{(3)}_4$ at two loops is a multiplicative order ${m\over n}$ correction and vanishes because the leading order
anomalous dimension vanishes. 
Evaluating the subleading order contribution coming from $D^{(2)}_4$, we find
\bea
   D_4^{(2)}\ket{i,R,r}=0\qquad i=5,6,7,8,9
\eea
so that the states that are BPS at the leading order do not receive a subleading correction. 
This is not peculiar to the example we consider and is to be expected generally, since for the BPS states we have
$n_{ab}(\sigma_2)=0$ for $a\ne b$. 

The state $\ket{4}$ also does not mix with other states.
However, for this state we have a nontrivial correction to the eigenvalue since
\bea
&&D_4^{(2)}\ket{4,R,r} = \sum_{T,t}\, 64\Bigg[ \delta_{RT}\left(\frac{\delta_{r'_1 t'_1}}{R_2}
+\frac{\delta_{r'_2 t'_2}}{R_1}\right)
\delta_{R''_{12} T''_{12}}(N+R_{1}-1)(N+R_{2}-2) \nonumber \\
&&\qquad + \delta_{RT}\frac{\delta_{r'_1 t'_1}}{R_{1}}\delta_{R''_{11}T''_{11}}
(N+R_{1}-1)(N+R_1-2)+ \delta_{RT}\frac{\delta_{r'_2 t'_2}}{R_{2}}
\delta_{R''_{22} T''_{22}}(N+R_{2}-2)(N+R_{2}-3) \nonumber \\
&&\qquad - \delta_{RT}\delta_{r'_1 t'_1}\delta_{R'_{1}T'_{1}}(N+R_{1}-1)
-\delta_{RT}\delta_{r'_2 t'_2}\delta_{R'_{2} T'_{2}}(N+R_{2}-2)\nonumber \\
&&\qquad - \delta_{R'_1T'_2} \frac{\delta_{r'_1 t'_2}}{\sqrt{R_{1}(R_{1} -1)}}
\delta_{R''_{11}T''_{12}}(N+R_{1}-2)\sqrt{(N+R_{1}-1)(N+R_{2}-1)}\nonumber \\
&&\qquad - \delta_{R'_2T'_1} \frac{\delta_{r'_2 t'_1}}
{\sqrt{R_{1}(R_{1} +1)}}\delta_{R''_{12}T''_{11}}(N+R_{1}-1)
\sqrt{(N+R_{1})(N+R_{2}-2)}\nonumber \\
&&\qquad - \delta_{R'_2T'_1} \frac{\delta_{r'_2 t'_1}}
{\sqrt{R_{2}(R_{2} -1)}}\delta_{R''_{22}T''_{12}}(N+R_{2}-3)
\sqrt{(N+R_{1})(N+R_{2}-2)}\nonumber \\
&&\qquad - \delta_{R'_1,T'_2} \frac{\delta_{r'_1 t'_2}}{\sqrt{R_{2}(R_{2} +1)}}
\delta_{R''_{12}T''_{22}}(N+R_{2}-2)\sqrt{(N+R_{1}-1)(N+R{2}-1)}\nonumber \\
&&\qquad + \delta_{R'_1 T'_2}\delta_{r'_1 t'_2}
\sqrt{(N+R_{1}-1)(N+R_{2}-1)}
+ \delta_{r'_2 t'_1}
\delta_{R'_{2} T'_{1}}\sqrt{(N+R_{1})(N+R_{2}-2)}
\Bigg]\ket{4,T,t}\cr
&& \label{D444}
\eea

The remaining states, $\ket{i}$ with $i=1,2,3$ mix under the action of $D_4^{(2)}$.
Using a matrix notation
\bea
D_4^{(2)}\ket{i,R,r}=\sum_{T,t}\, ({\cal D}_4^{(2)})_{ij}\ket{j,T,t}\qquad i,j=1,2,3
\eea 
the action of $D_4^{(2)}$ in this subspace is given by
\bea
{\cal D}_4^{(2)}=
A\left[
\begin{array}{ccc}
8 & 0 & 0 \\ 
0 & 4 & 0 \\ 
0 & 0 & 8
\end{array} 
\right]
+B\left[
\begin{array}{ccc}
0 & 1 & 0 \\ 
1 & 0 & 1 \\ 
0 & 1 & 0
\end{array} 
\right]+C
\left[
\begin{array}{ccc}
0 & 1 & 0 \\ 
0 & 0 & 1 \\ 
0 & 0 & 0
\end{array} 
\right]+C^\dagger
\left[
\begin{array}{ccc}
0 & 0 & 0 \\ 
1 & 0 & 0 \\ 
0 & 1 & 0
\end{array} 
\right]
\eea
where the coefficients $A,B$ and $C$ are quoted in Appendix \ref{results}.
The coefficients $A,B$ and $C$ are operators that have a nontrivial action of the $R,r$ labels of the Gauss graph operators.
It is straightforward to check that the matrix coefficients of these operators do not commute and hence they are not 
simulataneously diagonalizable.
This implies that the action of the dilatation operator no longer factorizes into commuting actions on the $Z$ and $Y$ 
fields.
It is this failure of factorization that we were referring to when we talked about the general structure of the mixing problem.

\section{Conclusions}\label{discussion}

Our results have a number of interesting features that deserve comment.
In the ${m\over n}=0$ limit, the action of the dilatation operator factorizes into an action on the $Z$ fields and
an action on the $Y$ fields.
The subleading correction has spoiled this factorization of the dilatation operator.
This is rather natural: in the limit ${m\over n}=0$ we consider a giant graviton built with an infinite
number ($n=\infty$) of $Z$ fields, so that the backreaction of the magnons on the giant graviton can be neglected.
Without backreaction, we expect the dyamics of the giant is completely decoupled from the dynamics of the 
magnons and this is the root of the factorized action of the dilatation operator.
By adding the first correction, we are saying that $n$ is large but not infinite.
In this situation, although back reaction is small, it is not zero.
The magnons will now provide a small perturbation to the dyamics of the giant, the action of the dilatation operator
will no longer factorize into an action on the giant (i.e. on the $Z$s) times an action on the magnons (i.e. on the $Y$s). 

The subleading correction spoiled the factorization of the dilatation operator by introducing further
operator mixing.
Another interesting results of our analysis, is that the subleading corrections did not induce extra mixing for the
BPS operators.
Indeed, after accounting for the complete ${m\over n}$ correction to two loops, we found our 
BPS operators remain uncorrected and continue to have a vanishing anomalous dimension.
Although our computation is performed in a specific example, we argued that we expect this conclusion to be general
since for the BPS operators we have $n_{ab}(\sigma)=0$ for $a\ne b$.
Looking at the result (\ref{fnlsbl}), it is clear that vanishing $n_{ab}(\sigma)$ implies a vanishing action of $D_4^{(2)}$.

The form of the action of the dilatation operator implies that when the correction to the anomalous dimension is non-zero
it will depend on the length of the rows of the Young diagrams labeling the operator.
Since these lengths determine the angular momentum of the giants, they determine the radius to which the giants will expand.
This implies that the anomalous dimensions start to depend on the geometry of the giant graviton.

The dynamics of open strings on the worldvolume of a giant graviton is expected to give rise to a Yang-Mills theory
at low energy.
The lightest mode of the open strings attached to the giant becomes the gauge boson of the theory.
This suggests that within ${\cal N}=4$ super Yang-Mills theory, we should see classes of operators whose
dynamics is captured by a new emergent gauge theory.
The acronym emergent is particularly apt in this case because this new Yang-Mills theory will be local on a space
that is distinct from the space of the original spacetime of the ${\cal N}=4$ super Yang-Mills theory.
The gauge symmetry which determines the interactions of the theory is a local symmetry with the respect
to this new space and the time of the original spacetime.
The space of the emergent Yang-Mills theory is the worldvolume of the giant graviton, which itself is built from
the $Z$s.
So this space and an associated local gauge invariance is to emerge from the dynamics of the $Z$ matrices in
the large $N$ limit.
For the operators dual to giant gravitons studied in this article, it is natural to think that the magnons themselves
will become the gauge bosons.
Indeed, the allowed state space of the magnons is parametrized by a double coset.
The structure of this double coset is determined by the expected Gauss Law of the emergent gauge theory.
To really understand the mechanism behind this emergence it is important that we get a good handle on how
the magnons interact.
It is by studying these interactions that we may hope to recognize the Yang-Mills theory that must emerge.
In this article we have computed the first of these interactions and we have developed tools that allow us
to study these interactions in general.

\noindent
{\it Acknowledgements:}
This work is based upon research supported by the South African Research Chairs
Initiative of the Department of Science and Technology and National Research Foundation.
Any opinion, findings and conclusions or recommendations expressed in this material
are those of the authors and therefore the NRF and DST do not accept any liability
with regard thereto.

\begin{appendix}

\section{Double Coset Background}\label{DCoset}

The double coset ansatz was formulated in \cite{deMelloKoch:2012ck}, by diaganolizing the one loop dilatation operator.
In this section we will review those aspects of \cite{deMelloKoch:2012ck} that are crucial for our study.

At the most basic level, the double coset ansatz follows from the fact that there are two ways to decompose 
$V_p^{\otimes m }$.
To start, refine $V_p^{\otimes m }$ by the $U(1)^p$ charges measured by $E_{ii}$, as follows
\bea 
  V_p = \oplus_{ i=1}^p V_i 
\eea
The vector space $V_i$ is a one-dimensional space.
It is spanned by the eigenstate of $E_{ii}$ with eigenvalue one. 
Consequently if $v_i \in V_i$ we have
\bea 
&& E_{ii} v_j = \delta_{ij}  v_i \cr 
&& E_{ij} v_k = \delta_{jk} v_i 
\eea
In the restricted Schur polynomial construction of \cite{Koch:2011hb} for long rows,  a state in $V_i$ corresponds to a  $Y$-box 
in the $i$'th row. 
The $U(1)$ charges of a restricted Schur polynomial can be collected into the vector $\vec m $, which corresponds to a vector 
with $m_1$ copies of $v_1 $ , $m_2 $ copies of $v_2$ etc. 
\bea\label{defvbar}  
| \bar v , \vec{m}\rangle \equiv | 
v_1^{\otimes m_1 } \otimes v_2^{\otimes m_2 } \otimes \cdots  v_p^{\otimes m_p }
\rangle
\eea
A general state with these charges is given by acting with a permutation 
\bea 
| v_{\sigma } \rangle  \equiv \sigma | v_1^{\otimes m_1 } \otimes v_2^{\otimes m_2 } \otimes \cdots  v_p^{\otimes m_p }
\rangle
\eea
where 
\bea 
\sigma | v_{i_1} \otimes \cdots \otimes v_{i_p } \rangle 
=  | v_{i_{ \sigma (1)} }\otimes  \cdots \otimes v_{i_{ \sigma( p)  } } \rangle 
\eea
This description enjoys a symmetry under
\bea
   H=S_{m_1}\times S_{m_2}\times\cdots\times S_{m_p}
\eea
and as a consequence, not all $\sigma $ give independent vectors 
\bea 
| v_{ \sigma } \rangle = | v_{ \sigma \gamma } \rangle
\eea
if $ \gamma \in H $. 

The restricted Schur polynomials are organized by reduction multiplicities of $U(p)$ to $U(1)^p$, which are counted by the
Kotska numbers and resolved by the Gelfand-Tsetlin patterns.
It is possible to prove  the equality of Kotska numbers and the branching multiplicity of $S_m \rightarrow H$.
This is a very direct indication that there are two possible ways to organize the local operators of the theory.

We can  develop the steps above at the level of a basis for $ V_p^{\otimes m} $. 
In terms of the branching coefficients, defined by
\bea
{1\over |H|}\sum_{ \gamma \in H } \Gamma^{(s)}_{ik} (\gamma)
=\sum_\mu B^{s \rightarrow 1_H}_{ i \mu}  B^{s \rightarrow 1_H}_{ k \mu}
\eea
we have
\bea\label{u1pdec}  
 | \vec m  , s, \mu ; i \rangle \equiv  \sum_j B^{s \rightarrow 1_H}_{ j \mu}
 | v_{ s , i ,  j }\rangle =  \sum_j B^{s \rightarrow 1_H}_{ j \mu}  \sum_{ \sigma\in S_m } \Gamma^{(s)}_{ij} ( \sigma ) | v_{ \sigma} \rangle
\eea
The $\mu$ index is a multiplicity for the reduction of $S_m$ into $H$.
We also have
\bea 
\langle  \vec{m} , u, \nu ; j|  =  {  d_u\over m! |H|  }  \sum_{ \tau \in S_m } 
\langle \bar v ,\vec{m} | \tau^{-1} \Gamma^{(u)}_{j k } ( \tau ) B^{u\to 1_H}_{k\nu }  
\eea
which ensures the correct normalization
\bea 
\langle \vec{m}, u, \nu ; j |\vec{n} , s , \mu ; i\rangle = \delta_{\vec{m}\vec{n}} \delta_{us} \delta_{ji} 
\delta_{\mu\nu} 
\eea
Finally, the group-theoretic coefficients 
\bea 
C_{\mu_1 \mu_2 }^{s }  ( \tau )  = 
|H|  \sqrt { d_s  \over m! } 
\Gamma^{(s)}_{kl} ( \tau ) B^{s \rightarrow 1_H} _{ k \mu_1} B^{s \rightarrow 1_H} _{l \mu_2 } 
\eea 
provide an orthogonal   transformation between double coset elements labeled by $\sigma$ and the restricted
Schur polynomials labeled by an irreducible representation  $s\vdash m$ and a pair of multiplicities $\mu_1,\mu_2 $.

\section{How to compute traces}\label{howtotrace}

In this Appendix we will compute the traces needed to evaluate the action of $D_4^{(2)}$ in the Gauss graph basis.
The generic form of the trace we need to evaluate is
\bea
{\rm Tr}(Ap_{s\mu\nu}Bp_{u\gamma\delta})\label{tocompute}
\eea
with $A$ and $B$ any arbitrary product of the $E^{(A)}_{ij}$s.
If we are able to compute this trace, we are able to evaluate the action of any differential operator that does not change 
the number of $Z$ or $Y$ fields,  on the Gauss graph operator in the displaced corners approximation.
This therefore provides a general method to exploit the simplifications of the large $N$ limit, for this class of operators. 

The Fourier transform we want to consider maps between functions labeled by an irreducible representation $s$ and
a pair of multiplicity labels and functions that take values on the double coset $H\setminus S_m /H$.
We can choose a permutation $\sigma$ to represent each class of the coset $\big[\sigma\big]$.
The transform is then
\bea
  \tilde f(\big[\sigma\big] )=\sum_{s,\alpha,\beta} \Gamma_s(\sigma)_{ab}B_{a\alpha}^{s\to 1_H}B_{b\beta}^{s\to 1_H}
   f(s,\alpha,\beta)
\eea
For further details the reader is referred to \cite{deMelloKoch:2012ck}.  

\subsection{Projector transformed}

In this section we will Fourier transform the intertwining map used to define the restricted Schur polynomial. 
The projector that participates in the trace (\ref{tocompute}) can be expressed as
\bea
p_{s\mu\nu}=\sum_{ a}|\vec{m},s,\mu,a\rangle\langle \vec{m},s,\nu,a|
\eea
We will make use of the relations
\bea
  \langle \vec{m},t,\nu,j|
={d_t\over m! |H|}
\sum_{\tau\in S_m}\langle \bar{v},\vec{m}|\tau^{-1}\Gamma^{(t)}_{jk}(\tau)
B^{t\to 1_H}_{k\nu}
\eea
and
\bea
|\vec{m},s,\mu,i\rangle =\sum_{\sigma\in S_m}\Gamma^{(s)}_{il}(\sigma)B_{l\mu}^{s\to 1_H}\sigma|\bar{v},\vec{m}\rangle
\eea
as well as
\bea
  \langle \bar{v},\vec{m}|\sigma |\bar{v},\vec{m}\rangle = \sum_{\gamma\in H}\delta (\sigma\gamma )
\eea
and
\bea
\sum_{\nu}  B^{s\to 1_H}_{c\nu} B^{s\to 1_H}_{d\nu}
={1\over |H|}\sum_{\gamma\in H}\Gamma^{(s)}(\gamma)_{cd}
\eea
We will use the notation $|v_\sigma\rangle=\sigma|\bar{v}\rangle$.
It is then simple to show that 
\bea
\sum_{s\,\mu\,\nu\, l\, m}p_{s\mu\nu}B^{s\to 1_H}_{m\mu}B^{s\to 1_H}_{l\nu}\Gamma^{(s)}_{ml}(\sigma)
&=&\sum_{s\,\mu\,\nu\, l\, m\, a}|\vec{m},s,\mu,a\rangle\langle \vec{m},s,\nu,a| B^{s\to 1_H}_{m\mu} 
B^{s\to 1_H}_{l\nu}\Gamma^{(s)}_{ml}(\sigma)\cr
&=&{1\over |H|^3}
\sum_{\psi\,\tau\in S_m}\sum_{\gamma_1,\gamma_2\in H}
|v_\psi\rangle\langle v_\tau |\delta (\tau^{-1}\psi\gamma_2 \sigma\gamma_1)
\eea
This last equation implies that the permutation applied to the ket and the permutation applied to the bra are related by
multiplication by a permutation representing the double coset element.

\subsection{Summing over $H$}

We consider an intertwining map $p_{s\mu\nu}$ built on the state $|\bar{v}_1,\vec{m}_1\rangle$ with symmetry group $H_1$ 
and intertwining map $p_{t\gamma\delta}$ built on the state $|\bar{v}_2,\vec{m}_2\rangle$ with symmetry group $H_2$.
We make no assumptions about $H_1$ and $H_2$.
In general they will be different groups and hence we Fourier transform $p_{s\mu\nu}$ and $p_{t\gamma\delta}$ to different
double cosets.
Using the result obtained above, the Fourier transform of 
\bea
{\rm Tr}(Ap_{s\mu\nu}Bp_{u\gamma\delta})
\eea
is
\bea
T&=&{1\over |H_1|^3|H_2|^3}\sum_{\gamma_i\in H_i}
\sum_{\tau_i\in H_i}
\sum_{\psi_i\in S_m}
\langle v_2|\gamma_2 \sigma_2^{-1}\tau_2\psi_2^{-1}A\psi_1|v_1\rangle
\langle v_1|\gamma_1 \sigma_1^{-1}\tau_1\psi_1^{-1}B\psi_2|v_2\rangle\cr
&=&{1\over |H_1|^3|H_2|^3}\sum_{\gamma_i\in H_i}
\sum_{\tau_i\in H_i}
\sum_{\psi_i\in S_m}
\langle v_2|\gamma_2 \sigma_2^{-1}\tau_2\psi_2^{-1}A\psi_1|v_1\rangle
\langle v_2|\psi_2^{-1}B^T \psi_1 \tau_1\sigma_1\gamma_1 |v_1\rangle\cr
&&
\eea
To get to the last line, we used the fact that the matrix element 
$\langle v_1|\gamma_1 \sigma_1^{-1}\tau_1\psi_1^{-1}B\psi_2|v_2\rangle$ is a real number and the
permutations are represented by matrices with real elements.
To make the discussion concrete, it is useful make a specific choice for $A$ and $B$.
This will allow us to illustrate the argument in a very concrete setting.
In the end we will state the general result.
Choose, for example
\bea
A=E^{(1)}_{ki}E^{(2)}_{ij}\qquad B=E^{(1)}_{jl}E^{(2)}_{lk}
\eea
Using the facts that
\bea
   \gamma |v_1\rangle &=& |v_1\rangle \qquad\qquad \forall\gamma\in H_1\cr
   \beta |v_2\rangle &=& |v_2\rangle \qquad\qquad \forall\beta\in H_2\cr
    \psi^{-1}E^{(a)}_{qr}\psi &=& E^{\psi(a)}_{qr}\qquad\qquad \forall\psi\in S_m
\eea
we readily find
\bea
T&=&{1\over |H_1| |H_2|}
\sum_{\psi_i\in S_m}
\langle v_2|\sigma_2^{-1}E^{\psi_2 (1)}_{ki}E^{\psi_2 (2)}_{ij}\psi_1|v_1\rangle
\langle v_2| E^{\psi_2 (1)}_{lj}E^{\psi_2 (2)}_{kl}\psi_1\sigma_1|v_1\rangle\cr
&=&{1\over |H_1| |H_2|}
\sum_{\psi_i\in S_m}
\langle v_2|\sigma_2^{-1}\psi_2^{-1}A \psi_2 \psi_1|v_1\rangle
\langle v_2| \psi_2^{-1}B^T \psi_2\psi_1\sigma_1|v_1\rangle
\eea
where on the last line we have written the general result.
Our next task is to compute the sums over $\psi_1$ and $\psi_2$.

\subsection{Summing over $S_m$}

In this subsection we consider two increasing difficult examples before we state the general result.
The first example is closely related to the trace needed to obtain the one loop dilatation operator.
Since we know the result of this trace, this example is a nice test of our ideas.
The second example is a simple but non-trivial example which will illustrate how the general case
works.
In the following section we will quote the resut for the general case. 

\subsubsection{First Evaluation}

Choose $A=E^{(1)}_{ab}$ and $B=E^{(1)}_{ba}$ to get
\bea
T&=&{1\over |H_1||H_2|}\sum_{\psi_i\in S_m}
\langle v_2 |\sigma_2^{-1}E^{\psi_2 (1)}_{ab}\psi_1|v_1\rangle
\langle v_2|E^{\psi_2(1)}_{ab}\psi_1\sigma_1 |v_1\rangle\cr
&=&{1\over |H_1||H_2|}\sum_{\psi_i\in S_m}
\langle v_1 |\psi^{-1}_1 E^{\psi_2 (1)}_{ba} \sigma_2|v_2\rangle
\langle v_2|E^{\psi_2(1)}_{ab}\psi_1\sigma_1 |v_1\rangle
\eea
We must turn a $b$ vector in $|v_1\rangle$ into an $a$ vector (and possibly permute) to get $|v_2\rangle$.
Since the ordering of the slots in $|v_1\rangle$ and $|v_2\rangle$ is arbitrary, we can remove this possible permutation
by declaring
\bea\label{related}
  |v_2\rangle = E^{(1)}_{ab}|v_1\rangle
\qquad |v_1\rangle = E^{(1)}_{ba}|v_2\rangle
\eea
Thus (in the computation below we denote by $S_a^1$ ($S_a^2$) the set of all slots in $|v_1\rangle$ ($|v_2\rangle$) that are 
filled with an $a$ vector)
\bea
T&=&{1\over |H_1||H_2|}\sum_{\psi_i\in S_m}
\langle v_1 |\psi^{-1}_1 E^{\psi_2 (1)}_{ba} \sigma_2 E^{(1)}_{ab}|v_1\rangle
\langle v_1|E^{(1)}_{ba}E^{\psi_2(1)}_{ab}\psi_1\sigma_1 |v_1\rangle\cr
&=&{1\over |H_1||H_2|}\sum_{\psi_i\in S_m}
\langle v_1 |\psi^{-1}_1 \sigma_2 E^{\sigma_2 \psi_2 (1)}_{ba} E^{(1)}_{ab}|v_1\rangle
\langle v_1|E^{(1)}_{ba}E^{\psi_2(1)}_{ab}\psi_1\sigma_1 |v_1\rangle\cr
&=&{1\over |H_1||H_2|}\sum_{\psi_i\in S_m}\sum_{\gamma_i\in H_1}
\delta \Big(\gamma_1\psi_1^{-1}\sigma_2 (1,\sigma_2\psi_2 (1)\, )\Big)
\delta \Big(\gamma_2\sigma_1^{-1}\psi_1^{-1} (1,\psi_2 (1)\, )\Big)\cr
&&\times\sum_{x\in S_a^2}\delta (\sigma_2\psi_2(1),x)\sum_{y\in S_a^2}\delta (\psi_2(1),y)\cr
&=&{1\over |H_1||H_2|}\sum_{\psi_2\in S_m}\sum_{\gamma_i\in H_1}
\delta\Big(\gamma_1\sigma_1\gamma_2 (1,\psi_2(1))\sigma_2(1,\sigma_2\psi_2(1)\Big)\cr
&&\times \sum_{x\in S_a^2}\delta (\sigma_2\psi_2(1),x)\sum_{y\in S_a^2}\delta (\psi_2(1),y)
\eea
Now consider the final sum over $\psi_2$. 
$\psi_2 (1)$ is the start point of an oriented edge in Gauss graph $\sigma_2$ and $\sigma_2\psi_2 (1)$ is the end 
point of the same edge.
The delta functions on the last line ensure that both endpoints of this string are attached to node $a$ in the Gauss graph.
This is swapped with the edge labeled 1 (i.e. the edge in the first slot) and compared to $\sigma_1$.
According to (\ref{related}), the edge in the first slot of $|v_1\rangle$ is attached to node $b$.
Thus the above sum is ensuring that when a closed loop on node $a$ of  $\sigma_2$ is removed and reattached to
node $b$ of $\sigma_2$ we get $\sigma_1$.
The above sum is non-zero only when $\sigma_1$ and $\sigma_2$ are related in this way.
The deltas only fix $\psi(1)$, so summing over $S_m$ the remaining ``unfixed'' piece of $\psi_2$ gives $(m-1)!$.
The first delta will, as usual, give the norm of Gauss graph $\sigma_1$ and we will get a non-zero contribution whenever
$\psi_2(1)$ is one of the values in $S_a^2$. There are $n_{aa}(\sigma_2)$ possible values.
Thus, when $T$ is non-zero it takes the value
\bea
T=(m-1)! n_{aa} (\sigma_2)|O(\sigma_1)|^2\label{frstans}
\eea
where we have assumed that both $\sigma_1$ and $\sigma_2$ have a total of $p$ nodes and we denote the number of oriented 
line segments stretching from node $k$ to node $l$ of $\sigma$ by $n_{kl}(\sigma)$.
We have denoted the ``norm'' of the Gauss graph operator by $|O(\sigma_1)|^2$.
This is the value of the two point function of the Gauss operator
\bea
|O(\sigma_1)|^2= \prod_{i=1}^p n_{ii}(\sigma_1)!\prod_{k,l=1, l\ne k}^p n_{kl}(\sigma_1)!
=\langle O_{R,r}(\sigma_1)^\dagger O_{R,r}(\sigma_1)\rangle
\eea
The value of the trace (\ref{frstans}) is in perfect agreement with the known result\cite{Koch:2014csa}.

\subsubsection{Second evaluation}

For the second example we consider, we choose
\bea
A=E^{(1)}_{ki}E^{(2)}_{ij}\qquad B=E^{(1)}_{jl}E^{(2)}_{lk}
\eea
There is some freedom in the placement of the indices on $A$ and $B$.
To see why this is the case, recall that we are evaluating the Fourier transform of
\bea
{\rm Tr}(Ap_{s\mu\nu}Bp_{u\gamma\delta})
\eea
The intertwining maps $p_{s\mu\nu}$ and $p_{u\gamma\delta}$ commute with any element of $S_m$.
Consequently we have
\bea
{\rm Tr}(Ap_{s\mu\nu}Bp_{u\gamma\delta})
={\rm Tr}(A\,\sigma\, \sigma^{-1}\, p_{s\mu\nu}Bp_{u\gamma\delta})
={\rm Tr}(A\, \sigma\, p_{s\mu\nu}\, \sigma^{-1}\, Bp_{u\gamma\delta})
\eea
where $\sigma$ is any element in $S_m$.
Choosing $\sigma=(12)$ and using the representation $(12)=E^{(1)}_{ab}E^{(2)}_{ba}$
where $a$ and $b$ are summed from 1 to $p$, as well as the product rule
\bea
E^{(A)}_{ab}E^{(B)}_{cd}=E^{(B)}_{cd}E^{(A)}_{ab}\qquad\qquad A\ne B \qquad\qquad\qquad
E^{(A)}_{ab}E^{(A)}_{cd}=\delta_{bc}E^{(A)}_{ad}\label{Ealg}
\eea
we find
\bea
A(12)=E^{(1)}_{ki}E^{(2)}_{ij} (12)=E^{(1)}_{ki}E^{(2)}_{ij}E^{(1)}_{ab}E^{(2)}_{ba}
=E^{(1)}_{kj}E^{(2)}_{ii}
\eea
\bea
(12)B=(12)E^{(1)}_{jl}E^{(2)}_{lk}=E^{(1)}_{ab}E^{(2)}_{ba}E^{(1)}_{jl}E^{(2)}_{lk}
=E^{(1)}_{ll}E^{(2)}_{jk}
\eea
This implies that we can rather consider $A=E^{(1)}_{kj}E^{(2)}_{ii}$ and $B=E^{(1)}_{ll}E^{(2)}_{jk}$
without changing the value of the trace.
In this case we have (argue as we did above and use $|v_2\rangle =E^{(1)}_{kj}|v_1\rangle$)
\bea
T
&=&{1\over |H_1||H_2|}\sum_{\psi_i\in S_m}
\langle v_2|\sigma_2^{-1}E^{\psi_2(1)}_{kj}E^{\psi(2)}_{ii}\psi_1|v_1\rangle
\langle v_2|E^{\psi_2(1)}_{ll}E^{\psi(2)}_{kj}\psi_1\sigma_1|v_1\rangle\cr
&=&{1\over |H_1||H_2|}\sum_{\psi_i\in S_m}
\langle v_1|\psi_1^{-1}E^{\psi_2(1)}_{jk}E^{\psi(2)}_{ii}\sigma_2|v_2\rangle
\langle v_2|E^{\psi_2(1)}_{ll}E^{\psi(2)}_{kj}\psi_1\sigma_1|v_1\rangle
\eea
Notice that when $E^{\psi(2)}_{ii}$ acts on $|v_2\rangle$, it does not change the identity of any of the vectors
appearing in $|v_2\rangle$. 
On the other hand, $E^{\psi_2(1)}_{jk}$ will turn an $e_k$ into an $e_j$.
Thus, again up to an arbitrary permutation which we can always remove, we must have
\bea
|v_2\rangle = E^{(1)}_{kj}|v_1\rangle
\eea
The trace now takes the value
\bea
T&=&{1\over |H_1||H_2|}\sum_{\psi_i\in S_m}
\langle v_1|\psi_1^{-1}E^{\psi_2(1)}_{jk}E^{\psi(2)}_{ii}\sigma_2 E^{(1)}_{kj}|v_1\rangle
\langle v_1|E^{(1)}_{jk}E^{\psi_2(1)}_{ll}E^{\psi(2)}_{kj}\psi_1\sigma_1|v_1\rangle\cr
&=&{1\over |H_1||H_2|}\sum_{\psi_i\in S_m}
\langle v_1|\psi_1^{-1}\sigma_2 E^{\sigma_2 \psi_2(1)}_{jk}E^{\sigma_2 \psi(2)}_{ii} E^{(1)}_{kj}|v_1\rangle
\langle v_1|E^{(1)}_{jk}E^{\psi_2(1)}_{ll}E^{\psi(2)}_{kj}\psi_1\sigma_1|v_1\rangle\cr
&=&{1\over |H_1||H_2|}\sum_{\psi_i\in S_m}\sum_{\gamma_i\in H_1}
\delta\Big(\gamma_1\psi_1^{-1}\sigma_2 (\sigma_2\psi_2 (1),1)\Big)
\delta\Big(\gamma_2\sigma_1^{-1}\psi_1^{-1} (\psi_2 (2),1)\Big)\cr
&&\times \sum_{x\in S^2_l}\delta (x,\psi_2(1))\sum_{y\in S^2_k}\delta (y,\psi_2(2))
\sum_{w\in S^2_k}\delta (w,\sigma_2\psi_2 (1))\sum_{v\in S^2_i}\delta (v,\sigma_2\psi_2(2))\cr
&=&{1\over |H_1||H_2|}\sum_{\psi_i\in S_m}\sum_{\gamma_i\in H_1}
\delta\Big(\gamma_1\sigma_1\gamma_2 (1,\psi_2(2))\sigma_2 (\sigma_2\psi_2 (1),1)\Big)\cr
&&\times \sum_{x\in S^2_l}\delta (x,\psi_2(1))\sum_{y\in S^2_k}\delta (y,\psi_2(2))
\sum_{w\in S^2_k}\delta (w,\sigma_2\psi_2 (1))\sum_{v\in S^2_i}\delta (v,\sigma_2\psi_2(2))\cr
&=&{(m-2)!\over |H_1||H_2|}n_{lk}(\sigma_2)n_{ki}(\sigma_2)\prod_{q=1}^p n_{qq}(\sigma_1)!
\prod_{r,s=1,\, r\ne s}^p n_{rs}(\sigma_1)!
\eea
whenever it is non-zero.

\subsection{General Result}

Recall that both $T$ and $R$ have $p$ long rows or columns.
For the general result we consider
\bea
\sum_{s\,\mu\,\nu\, l\, m}
\sum_{u\,\gamma\,\delta\, n\, p}
{\rm Tr}(Ap_{s\mu\nu}Bp_{u\gamma\delta})
B^{s\to 1_H}_{m\mu}B^{s\to 1_H}_{l\nu}\Gamma^{(s)}_{ml}(\sigma_1 )
B^{u\to 1_H}_{p\gamma}B^{u\to 1_H}_{n\delta}\Gamma^{(u)}_{pn}(\sigma_2 )\cr
={1\over |H_1||H_2|}\sum_{\psi_i\in S_m}
\langle v_2|\sigma_2^{-1}\, (\psi_2^{-1}A\psi_2 )\, \psi_1 |v_1\rangle
\langle v_2|\,(\psi_2^{-1}B^T\psi_2)\,\psi_1\sigma_1 |v_1\rangle\label{sumtodo}
\eea
where $A$ and $B$ are each products of a collection of the $E^{(\alpha)}_{ab}$s with $1\le a,b\le p$ and $1\le\alpha\le m$.
We say that $E^{(\alpha)}_{ab}$ occupies slot $\alpha$.
The sum over $\psi_2$ sums over the possible choices for the 
slots into which we place the factors of $E^{(\alpha)}_{ab}$ in $A$ and $B$.
Thus, the specific slots chosen for the factors in $A$ and $B$ are arbitrary - we must simply respect the relative ordering
of factors in $A$ and $B$, i.e. factors sharing the same slot in one labeling share the same slot in all labelings.
The sum over $\psi_1$ ensures that the relative labeling of the vectors appearing in $|v_1\rangle$ and $|v_2\rangle$
is arbitrary. 
Thus, the specific labeling of the directed edges in the Gauss graph is arbitrary which ensures that the above sum is
indeed defined on the relevant double cosets to which $\sigma_1$ and $\sigma_2$ belong.
There are two pieces of information that we need to read from $\sigma_1$, $\sigma_2$, $A$ and $B$: 

\begin{itemize}

\item[1] When is the sum nonzero?

\item[2] What is the value of the sum?

\end{itemize}
It is simplest to begin with the second question first.
Towards this end, consider the expressions for $A$ and $B$.
After using the algebra for the $E^{(\alpha)}_{ab}$ if needed, we know that at most a single $E^{(\alpha)}_{ab}$ acts
per slot in both $A$ and $B$. 
By inserting factors of
\bea
   \sum_{a=1}^p E^{(\alpha)}_{aa}={\bf 1}\label{resolveone}
\eea
if necessary, we can ensure that the same set of occupied slots appears in $A$ and $B$. 
For concreteness, assume that $q$ slots are occupied in both.
Use $i_\alpha^r$ ($i_\alpha^c$) to denote the row (column) indices of the $E_{ab}$ in the $\alpha$th slot in $B$ and
use $j_\alpha^r$ ($j_\alpha^c$) to denote the row (column) indices of the $E_{ab}$ in the $\alpha$th slot in $A$.
Thanks to the lessons we have learned from the examples treated above, when the sum is non-zero it is given by
\bea
{1\over |H_1||H_2|}\sum_{\psi_i\in S_m}
\langle v_2|\sigma_2^{-1}\, (\psi_2^{-1}A\psi_2 )\, \psi_1 |v_1\rangle
\langle v_2|\,(\psi_2^{-1}B^T\psi_2)\,\psi_1\sigma_1 |v_1\rangle\cr
={(m-q)! |O(\sigma_1)|^2\over |H_1||H_2|}\prod_{\alpha=1}^q n_{i_\alpha^c j_\alpha^r}(\sigma_2)
\label{frstresi}
\eea
If any particular $n_{ij}(\sigma_2)$ appears more than once, each new factor in the product is to be reduced by 1.
For example, $n_{12}(\sigma_2)^3$ would be replaced by $n_{12}(\sigma_2)(n_{12}(\sigma_2)-1)(n_{12}(\sigma_2)-2)$.
By taking the transpose of  (\ref{frstresi}), the value of the sum is not changed because it is a real number.
However, the roles of $\sigma_1$ and $\sigma_2$, as well as of $A$ and $B$ are reversed on the LHS of (\ref{frstresi}).
Consequently, we must also have
\bea
{1\over |H_1||H_2|}\sum_{\psi_i\in S_m}
\langle v_2|\sigma_2^{-1}\, (\psi_2^{-1}A\psi_2 )\, \psi_1 |v_1\rangle
\langle v_2|\,(\psi_2^{-1}B^T\psi_2)\,\psi_1\sigma_1 |v_1\rangle\cr
={(m-q)! |O(\sigma_2)|^2\over |H_1||H_2|}\prod_{\alpha=1}^q n_{j_\alpha^c i_\alpha^r}(\sigma_1)
\label{secondresi}
\eea
The equality of (\ref{frstresi}) and (\ref{secondresi}) defines our delta function.
We find that $\delta_{AB}([\sigma_1][\sigma_2])=1$ if
\bea
{(m-q)! |O(\sigma_1)|^2\over |H_1||H_2|}\prod_{\alpha=1}^q n_{i_\alpha^c j_\alpha^r}(\sigma_2)
=
{(m-q)! |O(\sigma_2)|^2\over |H_1||H_2|}\prod_{\alpha=1}^q n_{j_\alpha^c i_\alpha^r}(\sigma_1)
\eea
and it is zero otherwise.

We will sketch how the general result is proved. 
First, even if $A$ and $B$ straddle $q<m$ slots, by using (\ref{resolveone}) we can always introduce further 
$E^{(\alpha)}$s so that all $m$ slots are straddled. 
Thus, without loss of generality we can now focus on the $q=m$ case.
In this case, it is easy to prove that if $\prod_{\alpha=1}^q n_{i_\alpha^c j_\alpha^r}(\sigma_2)$ is non-zero, it is
given by
\bea
    \prod_{\alpha=1}^q n_{i_\alpha^c j_\alpha^r}(\sigma_2) = |O(\sigma_2)|^2
\eea
which proves the result.
Similarly, if $\prod_{\alpha=1}^q n_{j_\alpha^c i_\alpha^r}(\sigma_1)$ is non-zero, it is equal to
\bea
   \prod_{\alpha=1}^q n_{j_\alpha^c i_\alpha^r}(\sigma_1)=|O(\sigma_1)|^2
\eea
which again proves the general result.

\subsection{Illustration of the General Result}

To illustrate the formula derived in the previous section consider computing the trace for the case that
\bea
  A=E^{(1)}_{31} E^{(2)}_{11} E^{(3)}_{32} E^{(4)}_{12} E^{(5)}_{32} E^{(6)}_{33}\cr
  B=E^{(1)}_{23} E^{(2)}_{23} E^{(3)}_{11} E^{(4)}_{13} E^{(5)}_{31} E^{(6)}_{23}
\eea
For our example we have $m=6$, $q=6$ and $p=3$, so that $(m-q)!=1$.
We choose $\sigma_1$ and $\sigma_2$ as illustrated below
\begin{align*}
\text{{\Large $\sigma_1$}} =\quad \raisebox{-.4\height}{\includegraphics[scale=.55]{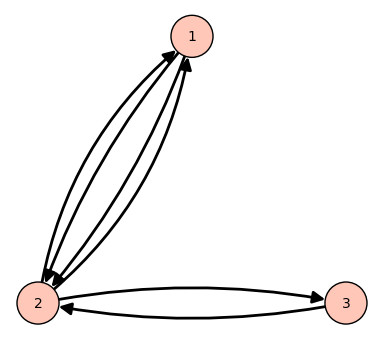}}\qquad\qquad
\text{{\Large $\sigma_2$}} =\quad \raisebox{-.4\height}{\includegraphics[scale=.55]{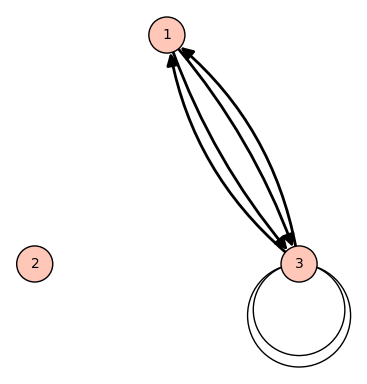}},
\end{align*}
From these Gauss graphs we easily read off $H_1=S_2\times S_3$ and $H_2=S_4\times S_2$.
Consequently $|H_1|=12$ and $|H_2|=48$.
If we choose the permutation
\bea
   \sigma_1=(13)(24)(56)\in H_1\setminus S_6/H_1
\eea
to represent the first Gauss graph, then we can choose the first factor in $H_1$ to permute 1 and 2 and the second factor  
to permute 3,4 and 5.
If we choose the permutation
\bea
    \sigma_2=(12)(45)
\eea
to represent the second Gauss graph, then we can choose the first factor to permute 2 and 5 and the last factor to 
permute 1, 3, 4 and 6.

From the row indices of $A$ given by the ordered set $\{3,1,3,1,3,3\}$, and the column indices of $B$ given by the
ordered set $\{3,3,1,3,1,3\}$, we read off
\bea
\prod_{\alpha=1}^q n_{i_\alpha^c j_\alpha^r}(\sigma_2)=
n_{33}(\sigma_2)n_{31}(\sigma_2)n_{13}(\sigma_2)n_{31}(\sigma_2)n_{13}(\sigma_2)n_{33}(\sigma_2)\cr
\to
n_{33}(\sigma_2)(n_{33}(\sigma_2)-1)
n_{31}(\sigma_2)(n_{31}(\sigma_2)-1)n_{13}(\sigma_2)(n_{13}(\sigma_2)-1)\cr
=|O(\sigma_2)|^2 = 8
\eea
which indicates that the sum may indeed be non-zero.
From the column indices of $A$ given by the ordered set $\{1,1,2,2,2,3\}$, and the row indices of $B$ given by the
ordered set $\{2,2,1,1,3,2\}$, we read off
\bea
\prod_{\alpha=1}^q n_{j_\alpha^c i_\alpha^r}(\sigma_1)=
n_{12}(\sigma_1)n_{12}(\sigma_1)n_{21}(\sigma_1)n_{21}(\sigma_1)n_{23}(\sigma_1)n_{32}(\sigma_1)\cr
\to
n_{12}(\sigma_1)(n_{12}(\sigma_1)-1)
n_{21}(\sigma_1)(n_{21}(\sigma_1)-1)n_{23}(\sigma_1)n_{32}(\sigma_1)\cr
=|O(\sigma_1)|^2 = 4
\eea
which indicates that the sum is indeed non-zero. We finally obtain
\bea
{1\over |H_1||H_2|}\sum_{\psi_i\in S_m}
\langle v_2|\sigma_2^{-1}\, (\psi_2^{-1}A\psi_2 )\, \psi_1 |v_1\rangle
\langle v_2|\,(\psi_2^{-1}B^T\psi_2)\,\psi_1\sigma_1 |v_1\rangle
={32\over |H_1||H_2|} ={1\over 18}
\eea

\subsection{Results for the 2-brane 4-string system}\label{results}

In this section we collect the operator valued coefficients that are relevant for the example described in section
\ref{2branes4strings}. The coefficients are:
\bea
A&=&
\delta_{RT}\left(\frac{\delta_{r'_1 t'_1}}{R_{2}}+\frac{\delta_{r'_2 t'_2}}{R_{1}}\right)
\delta_{R''_{12}T''_{12}}(N+R_{1}-1)(N+R_{2}-2)\cr
&+&\delta_{RT}\frac{\delta_{r'_1 t'_1}}{R_{1}}\delta_{R''_{11} T''_{11}}(N+R_{1}-1)(N+R_{1}-2)\cr
&+&\delta_{RT}\frac{\delta_{r'_2 t'_2}}{R_{2}}\delta_{R''_{22}T''_{22}}(N+R_{2}-2)(N+R_{2}-3)\cr
&-&2\delta_{RT}\delta_{r'_1 t'_1}\delta_{R'_{1} T'_{1}}(N+R_{1}-1)
-2\delta_{RT}\delta_{r'_2 t'_2}\delta_{R'_{2} T'_{2}}(N+R_{2}-2)\cr
&-&\delta_{R'_1T'_2}\frac{\delta_{r'_1 t'_2}}{\sqrt{R_{1}(R_{1}-1)}}
\delta_{R''_{11}T''_{12}}(N+R_{1}-2)\sqrt{(N+R_{1}-1)(N+R_{2}-1)}\cr
&-&\delta_{R'_2 T'_1}\frac{\delta_{r'_2t'_1}}{\sqrt{R_{1}(R_{1}+1)}}
\delta_{R''_{12}T''_{11}}(N+R_{1}-1)\sqrt{(N+R_{1})(N+R_{2}-2)}\cr
&-&\delta_{R'_2 T'_1}\frac{\delta_{r'_2 t'_1}}{\sqrt{R_{2}(R_{2}-1)}}
\delta_{R''_{22}T''_{12}}(N+R_{2}-3)\sqrt{(N+R_{1})(N+R_{2}-2)}\cr
&-&\delta_{R'_1 T'_2}\frac{\delta_{r'_1t'_2}}{\sqrt{R_{2}(R_{2}+1)}}\delta_{R''_{12}T''_{22}}
(N+R_{2}-2)\sqrt{(N+R_{1}-1)(N+R_{2}-1)}\cr
&+&2\delta_{r'_2t'_1}\delta_{R'_{2}T'_{1}}\sqrt{(N+R_{1})(N+R_{2}-2)}
+2\delta_{r'_1t'_2}\delta_{R'_{1}T'_{2}}\sqrt{(N+R_{1}-1)(N+R_{2}-1)}\cr
&&
\eea
\bea
B&=&
-8\delta_{r'_2t'_1}\delta_{RT}\delta_{R''_{12}T''_{12}}\sqrt{\frac{(N+R_{1}-1)(N+R_{2}-1)}{R_{1}R_{2}}}
\eea
\bea
C&=&8\frac{\delta_{r'_1t'_2}}{\sqrt{R_1(R_2+2)}}\delta_{R''_{11}T''_{22}}
\sqrt{(N+R_1 - 1)(N+R_1-2)(N+R_2-1)(N+R_2)}\cr
&-&8\delta_{R'_1 T'_2}\frac{\delta_{r'_1t'_1}}{\sqrt{R_1(R_2+1)}}\delta_{R''_{11}T''_{12}}(N+R_1-2)
\sqrt{(N+R_1-1)(N+R_2-1)}\cr
&-&8\delta_{R'_1 T'_2}\frac{\delta_{r'_2t'_2}}{\sqrt{R_1(R_2+1)}}\delta_{R''_{12}T''_{22}}(N+R_2-2)
\sqrt{(N+R_1-1)(N+R_2-1)}
\eea
\bea
C^\dagger&=&
8\frac{\delta_{r'_{2}t'_{1}}}{\sqrt{R_2(R_1+2)}}
\delta_{R''_{22}T''_{11}}\sqrt{(N+R_1)(N+R_1+1)(N+R_2-2)(N+R_2-3)}\cr
&-&8\delta_{R'_2 T'_1}\frac{\delta_{r'_1t'_1}}{\sqrt{R_2,(R_1+1)}}
\delta_{R''_{12}T''_{11}}(N+R_1-1)\sqrt{(N+R_1)(N+R_2-2)}\cr
&-&8\delta_{R'_2 T'_1}\frac{\delta_{r'_2t'_2}}{\sqrt{R_2(R_1+1)}}
\delta_{R''_{22}T''_{12}}(N+R_2-3)\sqrt{(N+R_1)(N+R_2-2)}
\eea
\end{appendix}


\begin{thebibliography}{} 

\bibitem{malda} 
  J.~M.~Maldacena,
  ``The Large N limit of superconformal field theories and supergravity,''
  Adv.\ Theor.\ Math.\ Phys.\  {\bf 2} (1998) 231
   [Int.\ J.\ Theor.\ Phys.\  {\bf 38} (1999) 1113]
  [hep-th/9711200].
  %%CITATION = HEP-TH/9711200;%%

\bibitem{mst} 
  J.~McGreevy, L.~Susskind and N.~Toumbas,
  ``Invasion of the giant gravitons from Anti-de Sitter space,''
  JHEP {\bf 0006} (2000) 008
  [hep-th/0003075].
  %%CITATION = HEP-TH/0003075;%%

\bibitem{myers} 
  M.~T.~Grisaru, R.~C.~Myers and O.~Tafjord,
  ``SUSY and goliath,''
  JHEP {\bf 0008} (2000) 040
  [hep-th/0008015].
  %%CITATION = HEP-TH/0008015;%%

\bibitem{hash}
   A.~Hashimoto, S.~Hirano and N.~Itzhaki,
  ``Large branes in AdS and their field theory dual,''
  JHEP {\bf 0008} (2000) 051
  [hep-th/0008016].
  %%CITATION = HEP-TH/0008016;%%

\bibitem{bbns}
  V.~Balasubramanian, M.~Berkooz, A.~Naqvi and M.~J.~Strassler,
  ``Giant gravitons in conformal field theory,''
  JHEP {\bf 0204} (2002) 034
  [hep-th/0107119].
  %%CITATION = HEP-TH/0107119;%%

\bibitem{cjr1}
  S.~Corley, A.~Jevicki and S.~Ramgoolam,
  ``Exact correlators of giant gravitons from dual N=4 SYM theory,''
  Adv.\ Theor.\ Math.\ Phys.\  {\bf 5} (2002) 809
  [hep-th/0111222].
  %%CITATION = HEP-TH/0111222;%%

\bibitem{cjr2}
  S.~Corley and S.~Ramgoolam,
  ``Finite factorization equations and sum rules for BPS correlators in N=4 SYM theory,''
  Nucl.\ Phys.\ B {\bf 641}, 131 (2002)
  doi:10.1016/S0550-3213(02)00573-4
  [hep-th/0205221].
  %%CITATION = doi:10.1016/S0550-3213(02)00573-4;%%

\bibitem{dssi} 
  R.~de Mello Koch, J.~Smolic and M.~Smolic,
  ``Giant Gravitons - with Strings Attached (I),''
  JHEP {\bf 0706} (2007) 074
  [hep-th/0701066].
  %%CITATION = HEP-TH/0701066;%%

%\cite{Kimura:2007wy}
\bibitem{Kimura:2007wy} 
  Y.~Kimura and S.~Ramgoolam,
  ``Branes, anti-branes and brauer algebras in gauge-gravity duality,''
  JHEP {\bf 0711}, 078 (2007)
  doi:10.1088/1126-6708/2007/11/078
  [arXiv:0709.2158 [hep-th]].
  %%CITATION = doi:10.1088/1126-6708/2007/11/078;%%

\bibitem{BHR1}
  T.~W.~Brown, P.~J.~Heslop and S.~Ramgoolam,
  ``Diagonal multi-matrix correlators and BPS operators in N=4 SYM,''
  arXiv:0711.0176 [hep-th].
  %%CITATION = ARXIV:0711.0176;%%

\bibitem{BHR2}
  T.~W.~Brown, P.~J.~Heslop and S.~Ramgoolam,
  ``Diagonal free field matrix correlators, global symmetries and giant
  gravitons,''
  arXiv:0806.1911 [hep-th].
  %%CITATION = ARXIV:0806.1911;%%

%\cite{Bhattacharyya:2008rb}
\bibitem{Bhattacharyya:2008rb}
  R.~Bhattacharyya, S.~Collins and R.~d.~M.~Koch,
  ``Exact Multi-Matrix Correlators,''
  JHEP {\bf 0803}, 044 (2008)
  [arXiv:0801.2061 [hep-th]].
  %%CITATION = JHEPA,0803,044;%%

\bibitem{Kimura:2008ac} 
  Y.~Kimura and S.~Ramgoolam,
  ``Enhanced symmetries of gauge theory and resolving the spectrum of local operators,''
  Phys.\ Rev.\ D {\bf 78}, 126003 (2008)
  doi:10.1103/PhysRevD.78.126003
  [arXiv:0807.3696 [hep-th]].
  %%CITATION = doi:10.1103/PhysRevD.78.126003;%%

%\cite{Kimura:2009jf}
\bibitem{Kimura:2009jf} 
  Y.~Kimura,
  ``Non-holomorphic multi-matrix gauge invariant operators based on Brauer algebra,''
  JHEP {\bf 0912}, 044 (2009)
  doi:10.1088/1126-6708/2009/12/044
  [arXiv:0910.2170 [hep-th]].
  %%CITATION = doi:10.1088/1126-6708/2009/12/044;%%

%\cite{Kimura:2012hp}
\bibitem{Kimura:2012hp} 
  Y.~Kimura,
  ``Correlation functions and representation bases in free N=4 Super Yang-Mills,''
  Nucl.\ Phys.\ B {\bf 865}, 568 (2012)
  doi:10.1016/j.nuclphysb.2012.08.010
  [arXiv:1206.4844 [hep-th]].
  %%CITATION = doi:10.1016/j.nuclphysb.2012.08.010;%%

%\cite{Pasukonis:2013ts}
\bibitem{Pasukonis:2013ts} 
  J.~Pasukonis and S.~Ramgoolam,
  ``Quivers as Calculators: Counting, Correlators and Riemann Surfaces,''
  JHEP {\bf 1304}, 094 (2013)
  doi:10.1007/JHEP04(2013)094
  [arXiv:1301.1980 [hep-th]].
  %%CITATION = doi:10.1007/JHEP04(2013)094;%%

\bibitem{Balasubramanian:2002sa} 
  V.~Balasubramanian, M.~x.~Huang, T.~S.~Levi and A.~Naqvi,
  ``Open strings from N=4 superYang-Mills,''
  JHEP {\bf 0208}, 037 (2002)
  doi:10.1088/1126-6708/2002/08/037
  [hep-th/0204196].
  %%CITATION = doi:10.1088/1126-6708/2002/08/037;%%

%\cite{Berenstein:2003ah}
\bibitem{Berenstein:2003ah} 
  D.~Berenstein,
  ``Shape and holography: Studies of dual operators to giant gravitons,''
  Nucl.\ Phys.\ B {\bf 675}, 179 (2003)
  doi:10.1016/j.nuclphysb.2003.10.004
  [hep-th/0306090].
  %%CITATION = doi:10.1016/j.nuclphysb.2003.10.004;%%

\bibitem{Balasubramanian:2004nb} 
  V.~Balasubramanian, D.~Berenstein, B.~Feng and M.~x.~Huang,
  ``D-branes in Yang-Mills theory and emergent gauge symmetry,''
  JHEP {\bf 0503}, 006 (2005)
  doi:10.1088/1126-6708/2005/03/006
  [hep-th/0411205].
  %%CITATION = doi:10.1088/1126-6708/2005/03/006;%%

%\cite{Berenstein:2005fa,Berenstein:2006qk}
\bibitem{Berenstein:2005fa} 
  D.~Berenstein, D.~H.~Correa and S.~E.~Vazquez,
  ``Quantizing open spin chains with variable length: An Example from giant gravitons,''
  Phys.\ Rev.\ Lett.\  {\bf 95}, 191601 (2005)
  doi:10.1103/PhysRevLett.95.191601
  [hep-th/0502172].
  %%CITATION = doi:10.1103/PhysRevLett.95.191601;%%

%\cite{Berenstein:2006qk}
\bibitem{Berenstein:2006qk} 
  D.~Berenstein, D.~H.~Correa and S.~E.~Vazquez,
  ``A Study of open strings ending on giant gravitons, spin chains and integrability,''
  JHEP {\bf 0609}, 065 (2006)
  doi:10.1088/1126-6708/2006/09/065
  [hep-th/0604123].
  %%CITATION = doi:10.1088/1126-6708/2006/09/065;%%

\bibitem{dssii} 
  R.~de Mello Koch, J.~Smolic and M.~Smolic,
  ``Giant Gravitons - with Strings Attached (II),''
  JHEP {\bf 0709} (2007) 049
  [hep-th/0701067].
  %%CITATION = HEP-TH/0701067;%%

\bibitem{bds} 
  D.~Bekker, R.~de Mello Koch and M.~Stephanou,
  ``Giant Gravitons - with Strings Attached. III.,''
  JHEP {\bf 0802} (2008) 029
  [arXiv:0710.5372 [hep-th]].
  %%CITATION = ARXIV:0710.5372;%%

%\cite{Carlson:2011hy}
\bibitem{Carlson:2011hy}
  W.~Carlson, R.~d.~M.~Koch, H.~Lin,
  ``Nonplanar Integrability,''
  JHEP {\bf 1103}, 105 (2011).
  [arXiv:1101.5404 [hep-th]].
  %%CITATION = ARXIV:1101.5404;%%

%\cite{Koch:2011hb}
\bibitem{Koch:2011hb}
  R.~d.~M.~Koch, M.~Dessein, D.~Giataganas, C.~Mathwin,
 ``Giant Graviton Oscillators,'' [arXiv:1108.2761 [hep-th]].
  %%CITATION = ARXIV:1108.2761;%%

\bibitem{gs}
R.~de Mello Koch, G.~Kemp, S.~Smith, 
``From Large N Nonplanar Anomalous Dimensions to Open Spring Theory,'' [arXiv:1111.1058 [hep-th]]. 
  %%CITATION = ARXIV:1111.1058;%%

%\cite{deMelloKoch:2012ck}
\bibitem{deMelloKoch:2012ck} 
  R.~de Mello Koch and S.~Ramgoolam,
  ``A double coset ansatz for integrability in AdS/CFT,''
  JHEP {\bf 1206}, 083 (2012)
  [arXiv:1204.2153 [hep-th]].
  %%CITATION = ARXIV:1204.2153;%%

\bibitem{chrisnirina}
R.~de Mello Koch, N.~H.~Tahiridimbisoa and C.~Mathwin,
 ``Anomalous Dimensions of Heavy Operators from Magnon Energies,''
  arXiv:1506.05224 [hep-th].
  %%CITATION = ARXIV:1506.05224;%%

\bibitem{LLM}
  H.~Lin, O.~Lunin and J.~M.~Maldacena,
  ``Bubbling AdS space and 1/2 BPS geometries,''
  JHEP {\bf 0410}, 025 (2004)
  [hep-th/0409174].
  %%CITATION = HEP-TH/0409174;%%

\bibitem{db}
  D.~Berenstein, D.~H.~Correa and S.~E.~Vazquez,
  ``All loop BMN state energies from matrices,''
  JHEP {\bf 0602}, 048 (2006)
  [hep-th/0509015].
  %%CITATION = HEP-TH/0509015;%%

\bibitem{hm1}
  D.~M.~Hofman and J.~M.~Maldacena,
  ``Reflecting magnons,''
  JHEP {\bf 0711}, 063 (2007)
  [arXiv:0708.2272 [hep-th]].
  %%CITATION = ARXIV:0708.2272;%%

\bibitem{hm2}
  D.~M.~Hofman and J.~M.~Maldacena,
  ``Giant Magnons,''
  J.\ Phys.\ A {\bf 39}, 13095 (2006)
  [hep-th/0604135].
  %%CITATION = HEP-TH/0604135;%%

\bibitem{stuart}
R.~de Mello Koch, S.~Graham and I.~Messamah,
  ``Higher Loop Nonplanar Anomalous Dimensions from Symmetry,''
  JHEP {\bf 1402}, 125 (2014)
  [arXiv:1312.6227 [hep-th]].
  %%CITATION = ARXIV:1312.6227;%%

%\cite{Beisert:2005tm}
\bibitem{Beisert:2005tm} 
  N.~Beisert,
  ``The SU(2|2) dynamic S-matrix,''
  Adv.\ Theor.\ Math.\ Phys.\  {\bf 12}, 945 (2008)
  [hep-th/0511082].
  %%CITATION = HEP-TH/0511082;%%

\bibitem{twoloop}
R.~de Mello Koch, G.~Kemp, B.~A.~E.~Mohammed and S.~Smith,
``Nonplanar integrability at two loops,''
JHEP {\bf 1210}, 144 (2012)
[arXiv:1206.0813 [hep-th]].
  %%CITATION = ARXIV:1206.0813;%%

\bibitem{threeloop}
E.~Dzienkowski,
 ``Excited States of Open Strings From $\mathcal{N}=4$ SYM,''
  arXiv:1507.01595 [hep-th].

%\cite{Berenstein:2013md}
\bibitem{Berenstein:2013md} 
  D.~Berenstein,
  ``Giant gravitons: a collective coordinate approach,''
  Phys.\ Rev.\ D {\bf 87}, no. 12, 126009 (2013)
  [arXiv:1301.3519 [hep-th]].
  %%CITATION = ARXIV:1301.3519;%%

%\cite{Berenstein:2013eya}
\bibitem{Berenstein:2013eya} 
  D.~Berenstein and E.~Dzienkowski,
  ``Open spin chains for giant gravitons and relativity,''
  JHEP {\bf 1308}, 047 (2013)
  [arXiv:1305.2394 [hep-th]].
  %%CITATION = ARXIV:1305.2394;%%

%\cite{Berenstein:2014isa}
\bibitem{Berenstein:2014isa} 
  D.~Berenstein and E.~Dzienkowski,
  ``Giant gravitons and the emergence of geometric limits in beta-deformations of $ \mathcal{N}=4 $ SYM,''
  JHEP {\bf 1501}, 126 (2015)
  [arXiv:1408.3620 [hep-th]].
  %%CITATION = ARXIV:1408.3620;%%

%\cite{Berenstein:2014zxa}
\bibitem{Berenstein:2014zxa} 
  D.~Berenstein,
  ``On the central charge extension of the N=4 SYM spin chain,''
  arXiv:1411.5921 [hep-th].
  %%CITATION = ARXIV:1411.5921;%%

%\cite{Beisert:2003tq}
\bibitem{Beisert:2003tq} 
  N.~Beisert, C.~Kristjansen and M.~Staudacher,
  ``The Dilatation operator of conformal N=4 superYang-Mills theory,''
  Nucl.\ Phys.\ B {\bf 664}, 131 (2003)
  doi:10.1016/S0550-3213(03)00406-1
  [hep-th/0303060].
  %%CITATION = doi:10.1016/S0550-3213(03)00406-1;%%

%\cite{Koch:2014csa}
\bibitem{Koch:2014csa} 
  R.~de Mello Koch, R.~Kreyfelt and S.~Smith,
  ``Heavy Operators in Superconformal Chern-Simons Theory,''
  Phys.\ Rev.\ D {\bf 90}, no. 12, 126009 (2014)
  doi:10.1103/PhysRevD.90.126009
  [arXiv:1410.0874 [hep-th]].
  %%CITATION = doi:10.1103/PhysRevD.90.126009;%%

\end{thebibliography}
\end{document}